\documentclass[review]{elsarticle}

\usepackage{lineno}
\usepackage{xurl}
\usepackage{hyperref}
\usepackage{xcolor}
\usepackage{comment}
\hypersetup{breaklinks=true} 
\modulolinenumbers[5]

\journal{arXiv}

\biboptions{authoryear}

\begin{document}

\begin{frontmatter}

\title{Frequency and duration of low-wind-power events in Germany}

\author[mcc]{Nils Ohlendorf}
\ead{ohlendorf@mcc-berlin.net}

\author[diw,ethub]{Wolf-Peter Schill\corref{correspondingauthor}}
\cortext[correspondingauthor]{Corresponding author}\ead{wschill@diw.de}

\address[mcc]{Mercator Research Institute on Global Commons and Climate Change (MCC), EUREF Campus 19, Torgauer Straße 12-15, 10829 Berlin, Germany}
\address[diw]{German Institute for Economic Research (DIW Berlin), Mohrenstrasse 58, 10117 Berlin, Germany}
\address[ethub]{Energy Transition Hub, Climate \& Energy College, The University of Melbourne}

\begin{abstract}
In the transition to a renewable energy system, the occurrence of low-wind-power events receives increasing attention. We analyze the frequency and duration of such events for onshore wind power in Germany, based on 40 years of reanalysis data and open software. We find that low-wind-power events are less frequent in winter than in summer, but the maximum duration is distributed more evenly between months. While short events are frequent, very long events are much rarer. Every year, a period of around five consecutive days with an average wind capacity factor below 10\% occurs, and every ten years a respective period of nearly eight days. These durations decrease if only winter months are considered. The longest event in the data lasts nearly ten days. We conclude that public concerns about low-wind-power events in winter may be overrated, but recommend that modeling studies consider multiple weather years to properly account for such events.
\end{abstract}


\end{frontmatter}

\pagebreak

\section{Introduction\label{sec: Introduction}}

The Paris Agreement calls for an extensive decarbonization of the global economy. A major strategy for achieving this goal is a massive expansion of variable renewable energy sources, in particular solar photovoltaics (PV) and wind power \citep{de_coninck_2018}. While power generation from solar PV largely follows diurnal and seasonal cycles with annually repeating patterns, wind power is subject to more irregular inter-annual as well as intra-annual variations which are relevant from a security of supply perspective. In countries with growing shares of wind power, the occurrence of low-wind-power (LWP) events thus receives increasing attention.

This is particularly true in Germany. In the context of its energy transition, Germany is one of the global front-runners in wind power deployment. In 2018, a total capacity of 52.5 GW of onshore wind power was installed in Germany, generating 90.5 TWh of electricity. This corresponds to 15\% of German gross electricity consumption \citep{BMWi_2019}. Given the government’s targets to expand the share of renewables in electricity consumption to 65\% by 2030 and at least 80\% by 2050 \citep{Bundesregierung_2019}, the dependence of the German energy system on wind power is set to increase strongly in the future. Concerns about LWP events have been discussed in German media \citep{Welt_2017, Welt_2019} and in the German parliament \citep{Bundestag_Plenarprotokoll_2019}, and LWP events are also mentioned in the government's energy transition reporting \citep{Bundestag_Unterrichtung_2019}. In this context, the term \textit{Dunkelflaute} is increasingly used. It refers to a persistent situation with very low power generation from wind and solar PV, which would be especially challenging in the German winter season where PV availability is low and electric load has its peak. Yet no clear definition of this concept has been provided so far \citep{WD_2019}, and quantitative evidence on the frequency and duration of such events is missing. In Table~$15$ of \cite{Bundestag_Unterrichtung_2019}, an independent expert commission generally assumes a no-wind-no-solar period of two weeks.

Yet research on LWP events is sparse so far. In this paper, we contribute to filling this gap, focusing on onshore wind power in Germany. We provide an in-depth analysis of the frequency, duration, and magnitude of LWP events, making use of reanalysis data for 40 full years (1980 to 2019) and power curves of recently installed wind turbines. In doing so, we propose two definitions of LWP events and investigate three different thresholds of capacity factors (2\%, 5\% and 10\%). We also compare the spatial distributions of the most persistent LWP event and the mean electricity generation. Parts of our analysis explicitly focus on winter months: these are particularly relevant, as power generation from solar PV is relatively low during this season, while the German peak load also occurs in winter. In order to allow for the highest degree of transparency and reproducibility, we provide the source code of our analysis under a permissive open-source license \citep{Ohlendorf_2020}.

There are only few dedicated analyses on the frequency and duration of LWP events. Early contributions address reliability aspects of spatially dispersed wind power in California \citep{Kahn_1979} or in the midwestern United States \citep{Archer_2007}. Analyses explicitly focusing on LWP events only recently emerged. Yet these differ from our work, amongst other factors, with respect to geographical and temporal coverage, data sources used, and methodologies applied.
In particular, previous low-wind analyses mostly draw on local measurement data and either evaluate wind speeds \citep{Leahy_2013, Patlakas_2017} or wind power \citep{Handschy_2017, Kruyt_2017}. \cite{Leahy_2013} and \cite{Patlakas_2017} investigate low-wind events for Ireland and the North Sea area, respectively.
Both studies firstly evaluate low-wind events that are constantly below a given wind speed threshold, and secondly determine annual minimum moving average wind speeds for given durations, using extreme value distributions. \cite{Kruyt_2017} and \cite{Handschy_2017} go one step further and calculate respective power generation from wind speeds for Switzerland and the United States, using a power curve.
While the findings of these studies are necessarily idiosyncratic to the specific geographical applications, some common findings emerge. First, low-wind events are less frequent and less persistent if more, and spatially more dispersed, measurement stations are used. Second, there are generally less events in winter than in summer. 

The measurement-based analyses face challenges related to their data sources. In general, studies that draw on measured wind speeds are spatially biased, have low measurement densities, and extrapolation from measurement height to hub height is challenging because of distorting effects of terrain, elevations or buildings \citep{Sharp_2015}. Measurement data may further be subject to inconsistencies caused by changing equipment and measurement errors. Extreme event analyses further require consistent measurements over large time periods to sufficiently capture climatic variations.

These issues can be addressed by using long-term meteorological reanalysis data. Such data is increasingly applied for onshore wind energy modelling. Several studies focus on data accuracy and on validating models of wind power generation \citep{Decker_2012, Sharp_2015, Olauson_2015, Rose_2015, Staffell_2016, Gonzalez_2017, Germer_2019}. Other analyses deal with variability aspects of wind power, but do not focus on extreme low-wind events. For example, \cite{Grams_2017} explain longer-term fluctuations in European wind power generation with different types of weather regimes, based on MERRA-2 data. With similar approaches, \cite{Collins_2018} investigate inter-annual variations of European wind and solar power, and \cite{Santos_Alamillos_2017} explore optimal allocations of renewable generation capacity in a European super grid. For the contingent U.S. states, \cite{Shaner_2018} investigate the reliability of future power systems dominated by wind and/or solar PV, and \cite{Kumler_2019} explore inter-annual renewable variability for Texas. Yet none of these studies explicitly focuses on the frequency and duration of extreme low-wind-power events.

A notable reanalysis study that does focus on extreme wind events is conducted by \cite{Cannon_2015} for Great Britain. Using 33 years of MERRA as well as ERA-Interim data, the authors conclude that the frequency and duration of low-wind-power events can be approximated by a Poisson-like process. \cite{Weber_2019} also use ERA-Interim data for a superstatistical analysis of extreme wind power events at nine specific European sites, including one German onshore location. They find that the distribution of low-wind events has a heavy tail, as low-wind events may result from a combination of different weather and circulation patterns.\footnote{\cite{Weber_2019} base their analysis on wind speeds, not wind power generation, with a cut-off threshold of $4$~m/s.} In another analysis based on ERA-Interim reanalysis data and other sources, \cite{Raynaud_2018} define and investigate the occurrence of renewable ``energy droughts'', which are measured relative to average daily generation. They find that wind power droughts are both relatively frequent and relatively short in most European countries, compared to hydro power droughts.

We contribute to this emerging literature with a dedicated open-source, reanalysis-based study that investigates LWP events in Germany in detail. To the best of our knowledge, we are the first to use MERRA-2 data in this context, i.e.,~spatially and temporally consistent reanalysis data covering 40 years at 50 m above surface. Compared to  \cite{Cannon_2015}, we also make use of not only one, but three recent power curves to represent different types of wind turbines that are characteristic for different locations defined by mean wind speeds. Complementary to  \cite{Raynaud_2018}, we further present an alternative approach to defining and evaluating LWPs by looking either at hours that are constantly below a threshold, or at hours with a mean below a threshold.

\section{Methods and data\label{sec: Methods and data}}

\subsection{General approach\label{subsec: General approach}}
Based on wind speeds and power curves, we derive an hourly aggregated time series of capacity factors for wind power in Germany. First, we take wind speeds at 50 m above surface from the MERRA-2 reanalysis dataset, which covers 40 years from 1980 to 2019, and extrapolate to hub heights.\footnote{See Section~\ref{appendix_reanalysis} for further information on the use of reanalysis data for energy modelling.} Second, capacity factors of each MERRA-2 grid cell are calculated based on power curves of recently installed wind turbines. Third, we spatially aggregate these capacity factors using a weighting scheme that considers the current spatial distribution of onshore wind power capacity in Germany. Finally, we investigate the resulting time series of hourly aggregated capacity factors by applying a narrower and a wider definition of LWP events.

\subsection{Wind speeds derived from reanalysis data\label{subsec: Wind speeds}}

We use the MERRA-2
dataset provided by NASA \citep{Gelaro_2017}. Data is available starting from the year 1980. In contrast to several other global reanalysis datasets which have time resolutions of 3 to 6 hours and provide wind speeds at 10 m above surface, MERRA-2 includes hourly wind speed data at 50 m, which allows better modelling of wind power generation.

\begin{figure}[t]
\centering{} \includegraphics[width=0.6\textwidth]{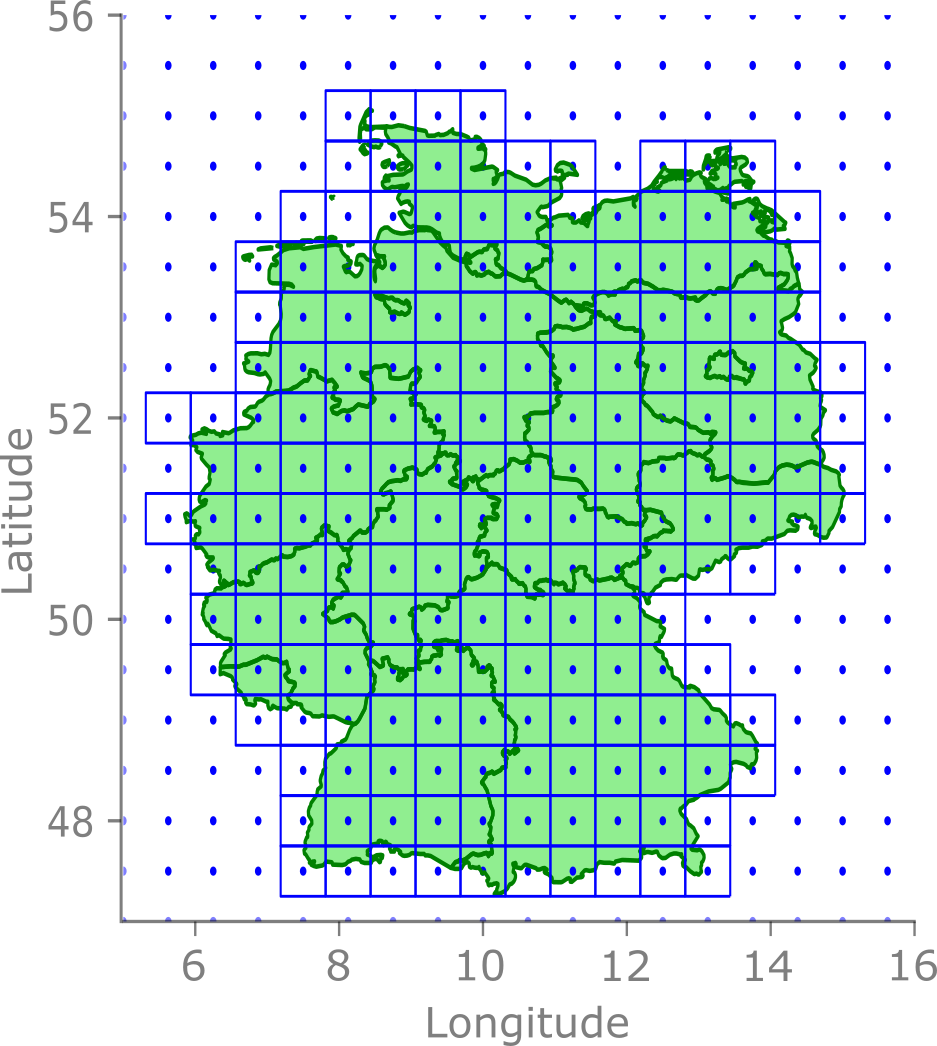}\\
\protect\caption{\label{fig: MERRA-2 grid points}MERRA-2 grid points (blue) and grid cells that intersect with Germany.}
\end{figure}

The MERRA-2 grid consists of 576 longitudinal and 361 latitudinal horizontal grid points, i.e.,~a resolution of $0.625^\circ$ x $0.5^\circ$ which for Germany roughly corresponds to 50 x 50 km \citep{Bosilovich_2016}. Figure~\ref{fig: MERRA-2 grid points} shows the grid points in blue and all grid cells extrapolated from these points that intersect with Germany. For each grid cell, MERRA-2 provides hourly northward and eastward wind speed data at 50 m above surface. Our dataset further includes surface roughness data for the year 2019.

\subsection{Aggregated wind power derived from wind speeds using power curves\label{subsec: Aggregated wind power}}

We calculate the magnitude of the horizontal wind speed ($U$) for each MERRA-2 grid point based on northward ($u$) and eastward components ($v$) at 50 m (Equation \ref{equ: 1}). 

\begin{equation}\label{equ: 1}
U=\sqrt{(u^2+v^2)}
\end{equation}

In line with \cite{Kruyt_2017}, we use the logarithmic power law to extrapolate wind speeds to hub-height ($h$) with $U_{hub}$ as the wind speed at hub height and $z_0$ as the surface roughness data for every grid point and each hour of the year 2019 (Equation \ref{equ: 2}).

\begin{equation}\label{equ: 2}
U_{hub}=w\frac{\ln\frac{h}{z_0}}{\ln\frac{50}{z_0}}
\end{equation}

\begin{figure}[t]
\centering{} \includegraphics[width=0.7\textwidth]{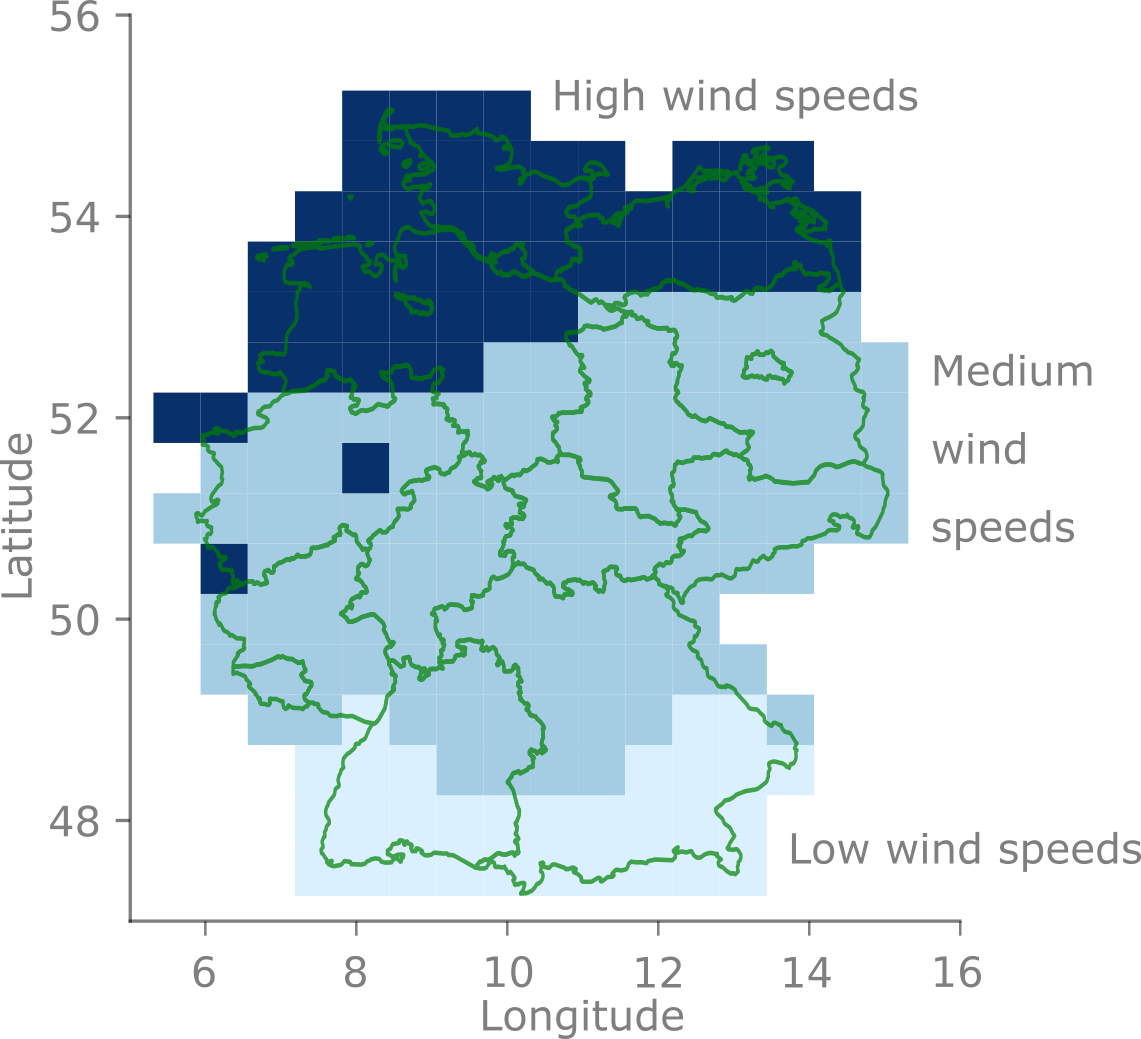}\\
\protect\caption{\label{fig: Wind speed zones}Wind speed zones in Germany. Dark blue implies high mean wind speeds, blue medium wind speeds, and light blue low mean wind speeds.}
\end{figure}

We define three types of wind zones, based on mean local wind speeds over 40 years for each MERRA-2 grid cell (Figure~\ref{fig: Wind speed zones}), and assign typical hub heights for wind turbines. For high-wind-speed sites, we assign a hub height of 100~m, for medium-wind-speed sites of 125~m, and for low-wind-speed sites of 139~m \citep{Deutsche_Windguard_2015}. These values reflect the mean hub heights of recently installed wind power plants in respective German wind speed zones. 

We calculate hourly capacity factors for each grid cell by applying power curves characteristic for the three wind zones. The power curves are based on manufacturer data of currently available wind turbines for low-, medium- and high-wind sites, respectively. Both the low- and high-wind site power curves represent an average of four wind turbines of similar diameters and capacities. We consider turbines from six manufacturers (see~\ref{sec: Wind_turbines}), among them four large companies which cover 87\% of the capacity installed in Germany in 2015 \citep{Deutsche_Windguard_2016}.

Manufacturers generally provide discrete capacity factors ($CF_{disc}$) for wind speed intervals of 1 m/s. For both the low- and high-wind curves, we first calculate discrete mean capacity factors for each wind speed and then calculate continuous capacity factors using a generalized logistic function (Equation \ref{equ: 3}).

\begin{equation}\label{equ: 3}
CF_{cont}=A+\frac{C}{(1+Te^{-B(U_{hub}-M)})^{1/T}}
\end{equation}

\begin{figure}[t]
\centering{} \includegraphics[width=0.9\textwidth]{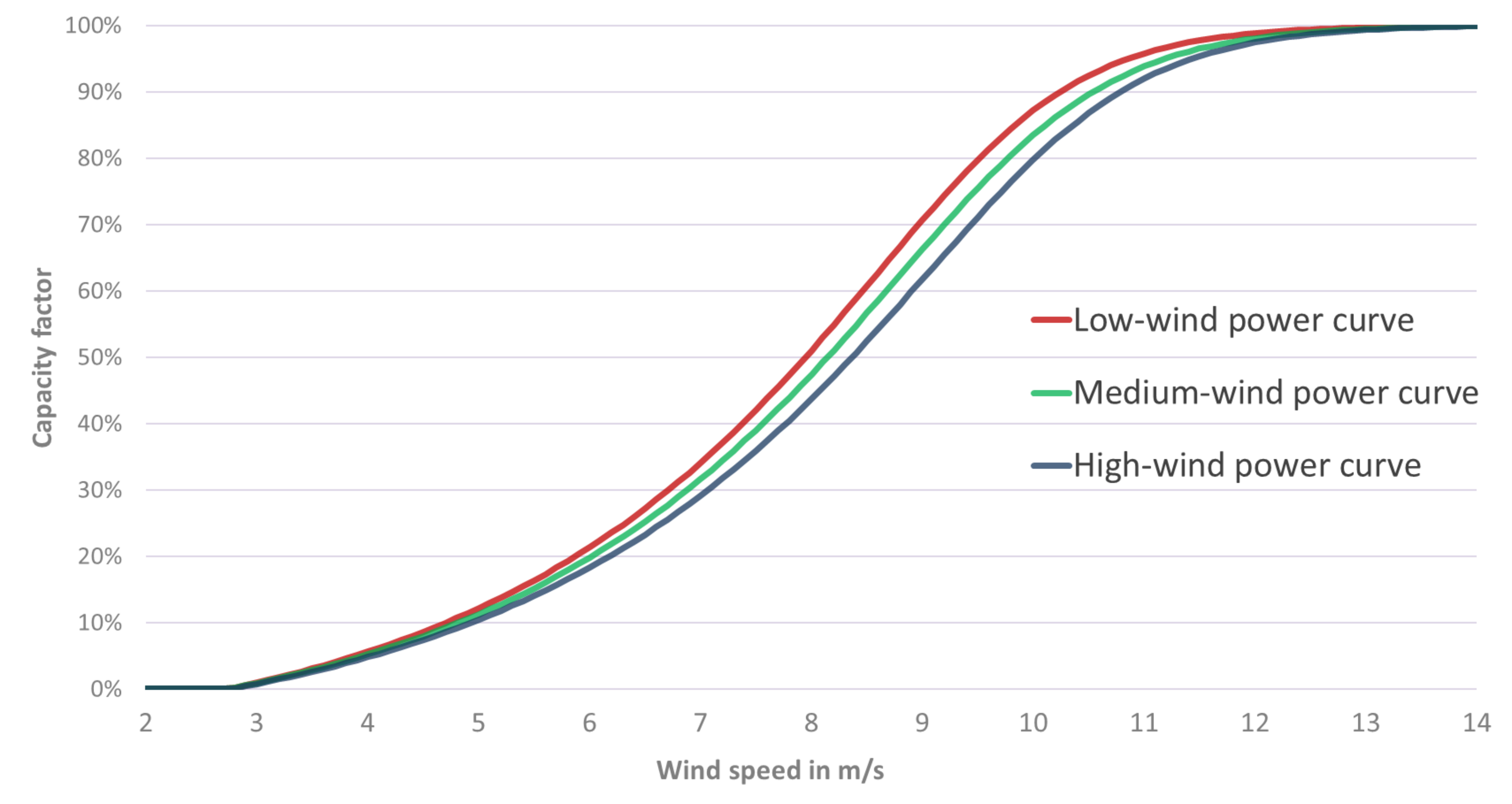}\\
\protect\caption{\label{fig: Power curves}Power curves of three types of wind turbines}
\end{figure}

Here, $CF_{cont}$ is the continuous capacity factor and $A$, $B$, $C$, $M$ and $T$ are fitted coefficients based on minimising the squared deviations between $CF_{disc}$ and $CF_{cont}$. For both the low- and the high-wind power curve, cut-in wind speeds of around 3~m/s emerge, and the resulting capacity factors are capped at 0\% and 100\%. The medium-wind power curve represents the average of the low- and high-wind curves (Figure~\ref{fig: Power curves}).  

\begin{figure}[t]
\centering{} \includegraphics[width=0.8\textwidth]{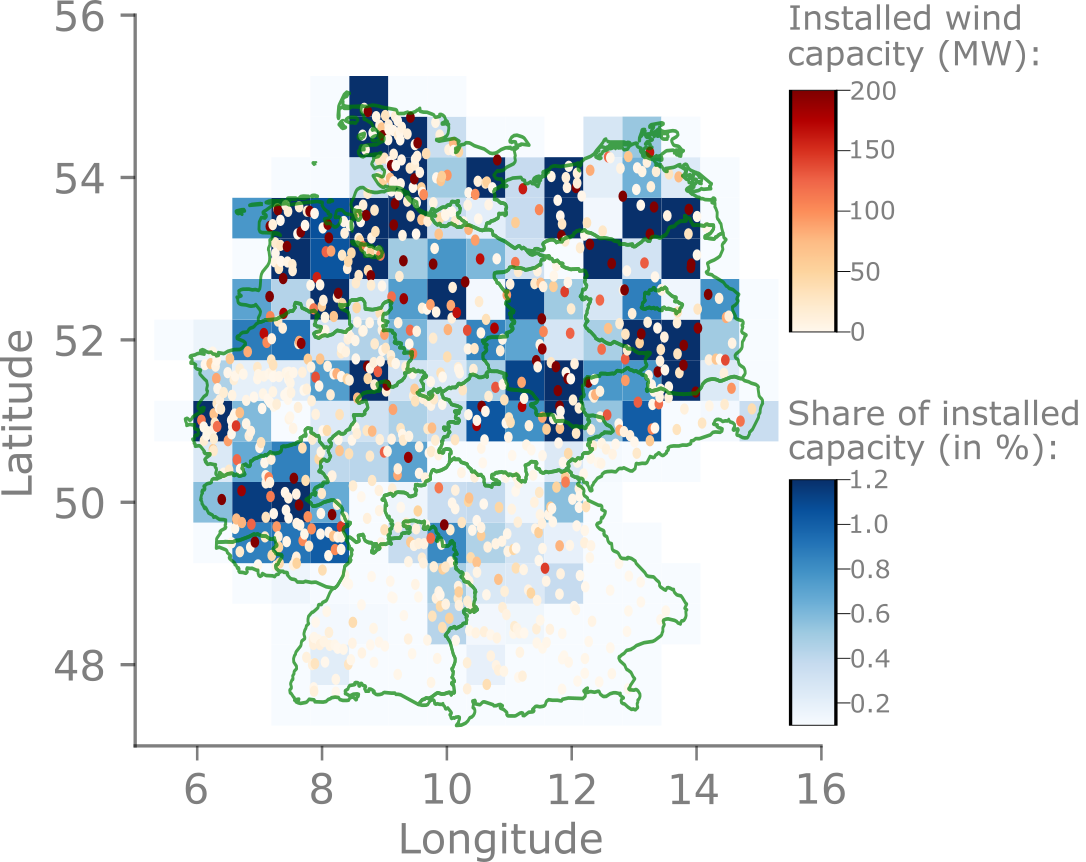}\\
\protect\caption{\label{fig: Wind power distribution}Distribution of currently installed wind power capacity in Germany. Darker colors indicate a larger share of total or relative installed capacity.}
\end{figure}

Aggregated hourly capacity factor time series for overall Germany are derived by weighting all grid cells with the current distribution of installed wind power generation capacity. The latter is extracted from Open Power System Data \citep{OPSD_2017, Wiese_2019} and open-source GIS data. The red points in Figure~\ref{fig: Wind power distribution} indicate the installed wind capacity of locally aggregated wind power plant sites in Germany and the blue squares show the corresponding relative capacity distribution of the MERRA-2 grid cells. Grid cells only partly intersecting with the German land area receive lower weights according to the overlapping area. We implicitly assume that the transmission infrastructure allows geographical balancing of wind power in Germany, which is currently largely the case.\footnote{This assumptions is particularly valid for low-wind periods. During high-wind, high-load periods, the German transmission grid can be constrained in North-South direction.}

\subsection{Definition of low-wind-power events\label{subsec: Definition of LWP events}}

\begin{figure}[t]
\centering{} \includegraphics[width=0.8\textwidth]{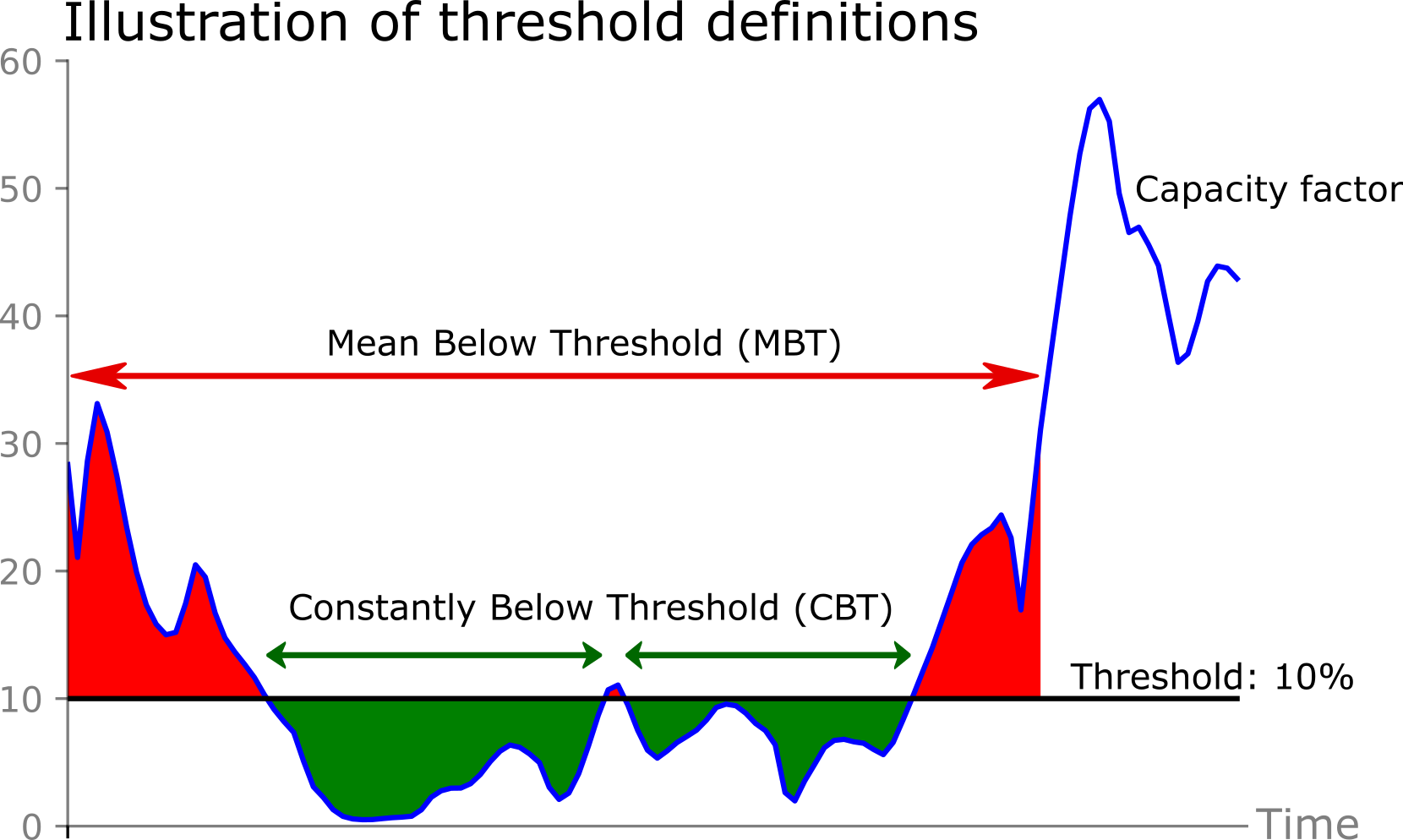}\\
\protect\caption{\label{fig: Definitions}Illustration of the two LWP event definitions}
\end{figure}

We propose two different measures of low-wind-power periods, a narrower and a wider one (Figure~\ref{fig: Definitions}). We further consider three alternative capacity factor thresholds of 2\%, 5\%, and 10\%.

As for the narrower definition, we consider LWP events to be consecutive hours in which the aggregated capacity factors are \textit{Constantly Below the Threshold} (CBT). This concept bears some resemblance to the ``runs analysis'' by \cite{Leahy_2013} or the ``duration given intensity'' method by \cite{Patlakas_2017}. Starting in the first hour, we list annual LWP events for durations starting from five consecutive hours and report the number of hours constantly below the given capacity factor threshold. We then increase the duration in hourly steps and repeat until there are no further events listed. 

To provide a wider definition, we consider LWP events to consist of consecutive hours in which the moving average of capacity factors is under the same threshold, i.e.,~\textit{Mean Below the Threshold} (MBT). Again, we list all LWP periods until we reach the threshold value, ensuring that LWP periods do not overlap. By definition, the MBT method results in more low-wind-power events for a given duration and also results in longer events for each threshold, compared to CBT.

The average annual amount of LWP events per duration over all 40~years equals the expected value of events per year. Further, the reciprocal value of the annual average provides the return period, that is the expected temporal distance between two similar reoccurring events. Periods overlapping annually or monthly are assigned to the year or month in which more than 50\% of the hours are located\footnote{
Accounting for annually overlapping periods requires December data from the previous year, and January data from the subsequent year. For the two boundary years 1980 and 2019, we substitute the missing data for December 1979 (January 2020) with data from December 1980 (January 2019).
}.

\section{Results\label{sec: Results}}

\subsection{Seasonal distribution and frequency of low-wind-power events\label{subsec: Seasonal distribution}}

\begin{figure}[h]
\centering{} \includegraphics[width=1\textwidth]{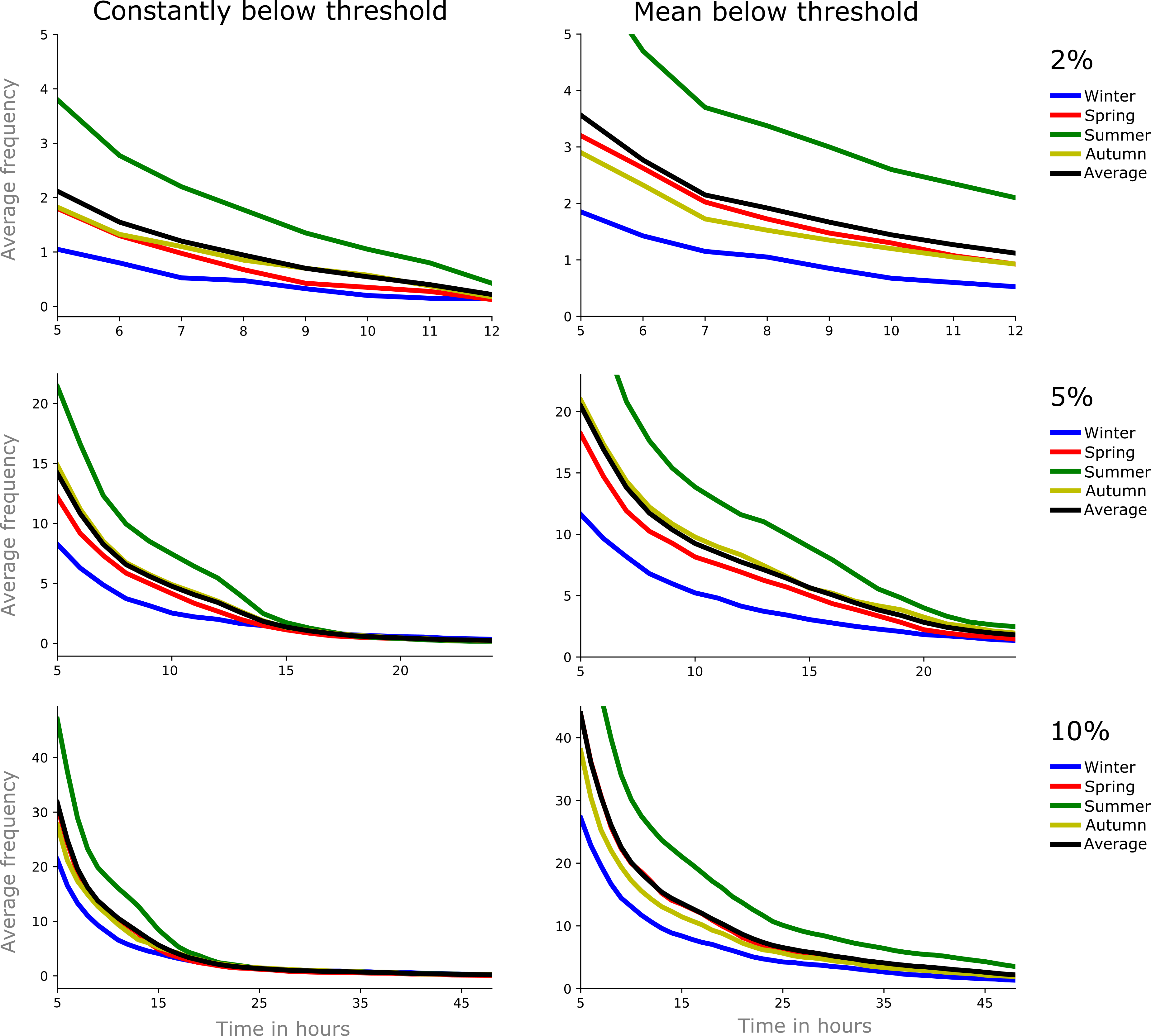}\\
\protect\caption{\label{fig: seasonal}Average seasonal duration (horizontal axis) and frequency (vertical axis) of LWP events in Germany}
\end{figure}

Figure~\ref{fig: seasonal} shows that LWP events are generally most frequent in summer (here defined as June-August) and least frequent in winter (December-February). The results for spring (March-May) and autumn (September-November) are mostly close to the annual average. Accordingly, respective findings made for other European countries \citep{Leahy_2013, Cannon_2015, Kruyt_2017} are also valid for Germany. 

The frequency of events for a given duration is about 1.5-3 times higher for the wider MBT definition compared to the narrower CBT concept. For both metrics, the frequency of LWP events increases substantially with the capacity factor threshold value. For example, a 10-hour event below a capacity factor of 2\% occurs on average around 0.2~times per winter for CBT and slightly less than once per winter for MBT. For a 10\% capacity factor threshold, there are on average around eight such winter events for CBT and 13 for MBT. In general, we find that short LWP events with a duration of up to around half a day are relatively frequent and may occur several times per year, especially under the wider MBT definition. Longer LWP events, in contrast, are much less frequent.

To provide a complementary perspective, we calculate the return periods for different durations of LWP events (Table \ref{tab: return periods}). The return periods are the reciprocal of the average (annual or seasonal) frequency of LWP events for different durations, considering both definitions and all three thresholds (cf. Figure~\ref{fig: seasonal}). For example, an LWP event with an average frequency of 0.2 for a given duration leads to a return period of 5 years for this specific duration. The longer a given duration, the lower its average frequency and the longer its return period.

For a return period of ten years, we find a duration of 17~hours (2\% capacity factor threshold), 41~hours (5\%) and 77~hours (10\%) under the narrower CBT definition, and a duration of 34~hours (2\%), 79~hours (5\%) and 188~hours (10\%) under the wider MBT concept. In other words, every ten years the German energy system has to deal with a period of nearly eight days of average wind power generation (MBT) below 10\% of the installed capacity.

\begin{table}\label{tab: return periods}
\protect\caption{Duration in hours for LWP events in winter or in any season for different return periods}
\centering
\tiny
\begin{tabular}{l|rrr|rrr|rrr|rrr}
~             & \multicolumn{6}{c|}{Constantly below threshold (CBT)}         & \multicolumn{6}{c}{Mean below threshold (MBT)}                \\ 
~             & \multicolumn{3}{c|}{Winter} & \multicolumn{3}{c|}{Any season} & \multicolumn{3}{c|}{Winter} & \multicolumn{3}{c}{Any season}  \\
Return period & 2\% & 5\% & 10\%            & 2\% & 5\% & 10\%                & 2\% & 5\% & 10\%            & 2\% & 5\% & 10\%                 \\ 
\hline
1 year        & 5   & 15  & 29              & 11  & 23  & 45                  & 8   & 30  & 63              & 18  & 58  & 122                  \\
2 years       & 7   & 21  & 40              & 13  & 32  & 57                  & 12  & 45  & 92              & 21  & 69  & 144                  \\
3 years       & 8   & 23  & 44              & 14  & 33  & 60                  & 14  & 52  & 101             & 23  & 71  & 161                  \\
4 years       & 9   & 30  & 48              & 14  & 33  & 63                  & 16  & 62  & 112             & 27  & 72  & 173                  \\
5 years       & 10  & 32  & 57              & 15  & 35  & 65                  & 22  & 68  & 113             & 28  & 75  & 178                  \\
6 years       & 10  & 32  & 57              & 15  & 35  & 67                  & 25  & 69  & 114             & 29  & 76  & 182                  \\
7 years       & 12  & 33  & 60              & 15  & 36  & 67                  & 27  & 70  & 114             & 31  & 76  & 186                  \\
8 years       & 14  & 33  & 63              & 17  & 37  & 69                  & 28  & 72  & 117             & 33  & 79  & 186                  \\
9 years       & 14  & 33  & 63              & 17  & 37  & 69                  & 28  & 72  & 117             & 33  & 79  & 186                  \\
10 years      & 14  & 33  & 64              & 17  & 41  & 77                  & 28  & 72  & 126             & 34  & 79  & 188                  \\
15 years      & 17  & 36  & 67              & 18  & 41  & 77                  & 31  & 76  & 129             & 38  & 82  & 189                  \\
20 years      & 19  & 41  & 77              & 19  & 49  & 81                  & 34  & 79  & 131             & 45  & 89  & 221                  \\
25 years      & 19  & 41  & 77              & 19  & 49  & 81                  & 34  & 79  & 131             & 45  & 89  & 221                  \\
30 years      & 19  & 41  & 77              & 19  & 49  & 81                  & 34  & 79  & 131             & 45  & 89  & 221                 
\end{tabular}
\end{table}

To better interpret these return periods, we provide an example for the German onshore wind power capacity of 52.5~GW installed in 2018. For this wind turbine fleet, average power generation is expected to not exceed around five~GW, i.e.,~10\% of capacity, during a period of around five consecutive days every year (122 hours, MBT for 'Any Season' in \ref{tab: return periods}). Every ten years, this period increases to nearly eight days, and every twenty years to more than nine full days. Looking only at LWP events in winter, these durations decrease to less than three days every winter, less than five days every tenth winter, and around five and a half days every twentieth winter.
The remaining load has to be covered by other generators, energy storage or demand-side measures. However, wind power still contributes some generation capacity above the 10\% threshold during some of these hours, as indicated by much lower CBT return periods.

\subsection{Magnitude of the most extreme low-wind-power events\label{subsec: Magnitude of most extreme events}}

\begin{figure}[h]
\centering{}
\includegraphics[width=0.9\textwidth]{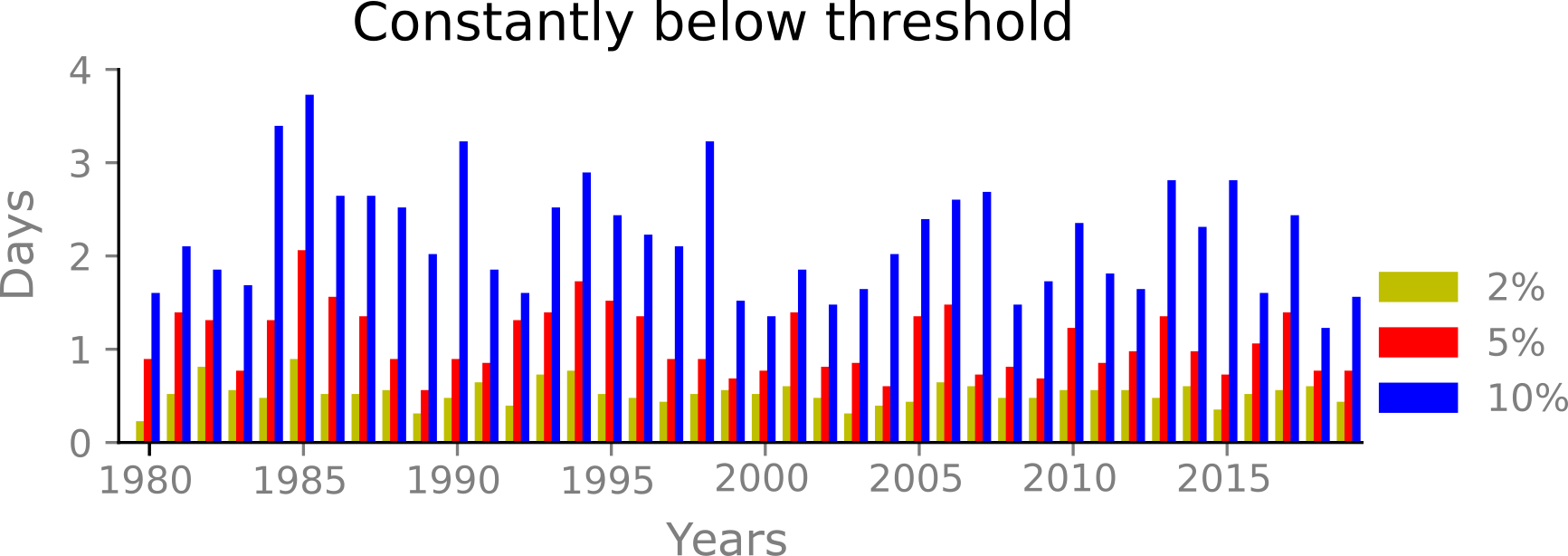}\\
\includegraphics[width=0.9\textwidth]{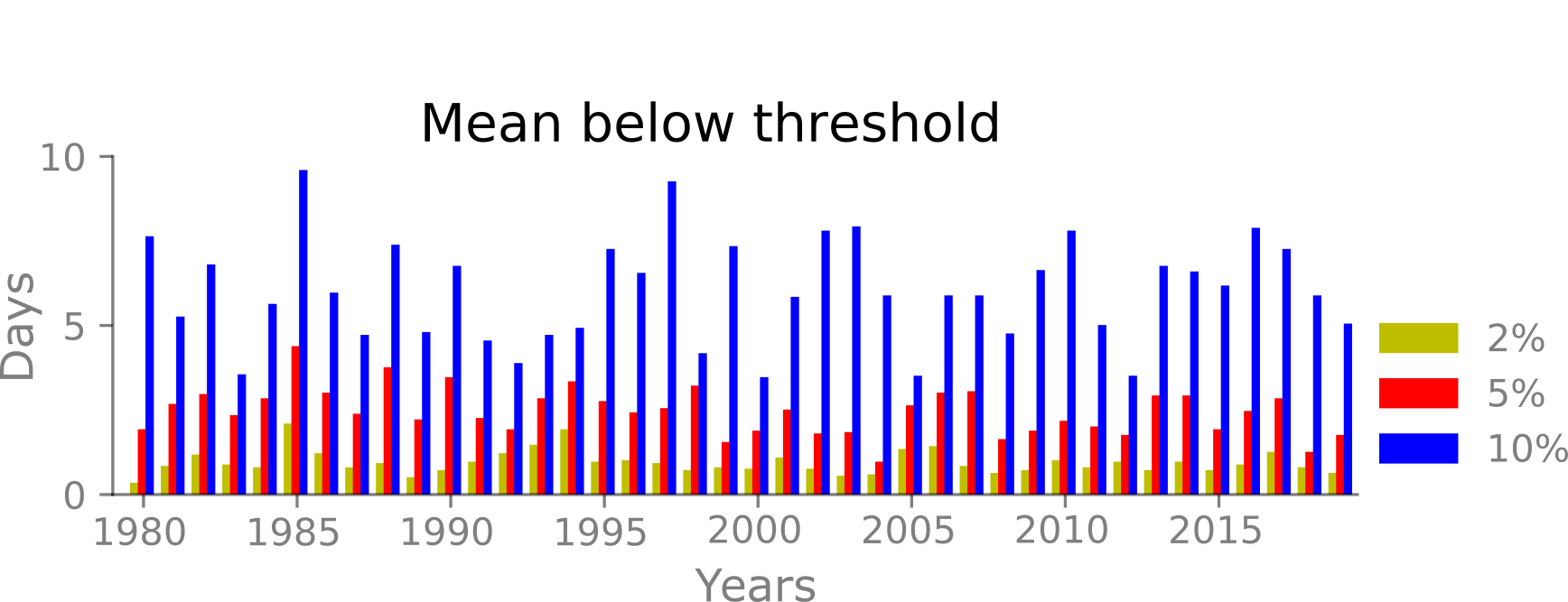}\\
\protect\caption{\label{fig: most extreme year}Most extreme LWP events per year. The vertical axis shows the duration of the longest event per year for the three capacity factor thresholds.}
\end{figure}

The most extreme LWP events over the entire 40 years analyzed can be interpreted as worst cases from an energy system planning perspective. In an annual perspective, the most extreme events occurred in 1985 for all capacity factor thresholds (Figure~\ref{fig: most extreme year}). Under the narrower CBT definition, there are nearly four consecutive days with wind power generation constantly below 10\% in 1985, and still around two consecutive days with generation constantly below 5\%. Under the wider MBT definition, the duration of this most extreme event increases to nearly ten days (10\%) or around four days (5\%).

While this 1985 event is the most extreme one under both CBT and MBT, the ranking of the second most extreme yearly events differs between the LWP definitions. For example, the second-longest event occurred in 1984 under the CBT definition. Yet under MBT, the duration of the most extreme event in 1984 is only average. In general, the definition of LWP events and the chosen thresholds have a substantial impact on quantitative results. Under MBT, the most extreme annual events are generally around twice as high compared to CBT.

We further find very large inter-annual variations. Considering the 10\% threshold, the longest event for the MBT definition lasted for almost 10 days in 1985, but in 2005 the longest duration was only three days for the same threshold. The relative difference between the longest events for each year increases with the threshold. These large variations of the most extreme annual LWP events complement the findings made by \cite{Collins_2018}, who determine large inter-annual variations of average renewable availability.

\begin{figure}[t]
\centering{}
\includegraphics[width=0.9\textwidth]{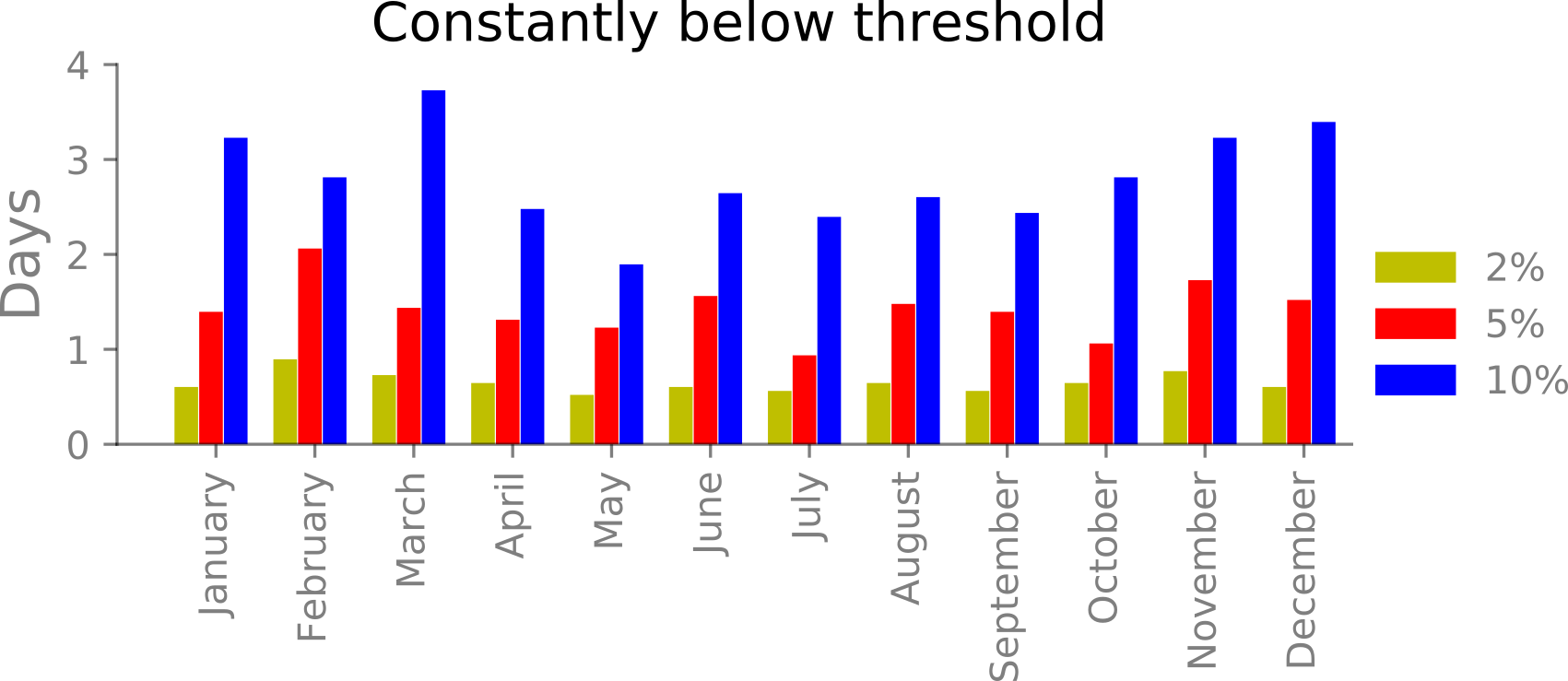}\\
\includegraphics[width=0.9\textwidth]{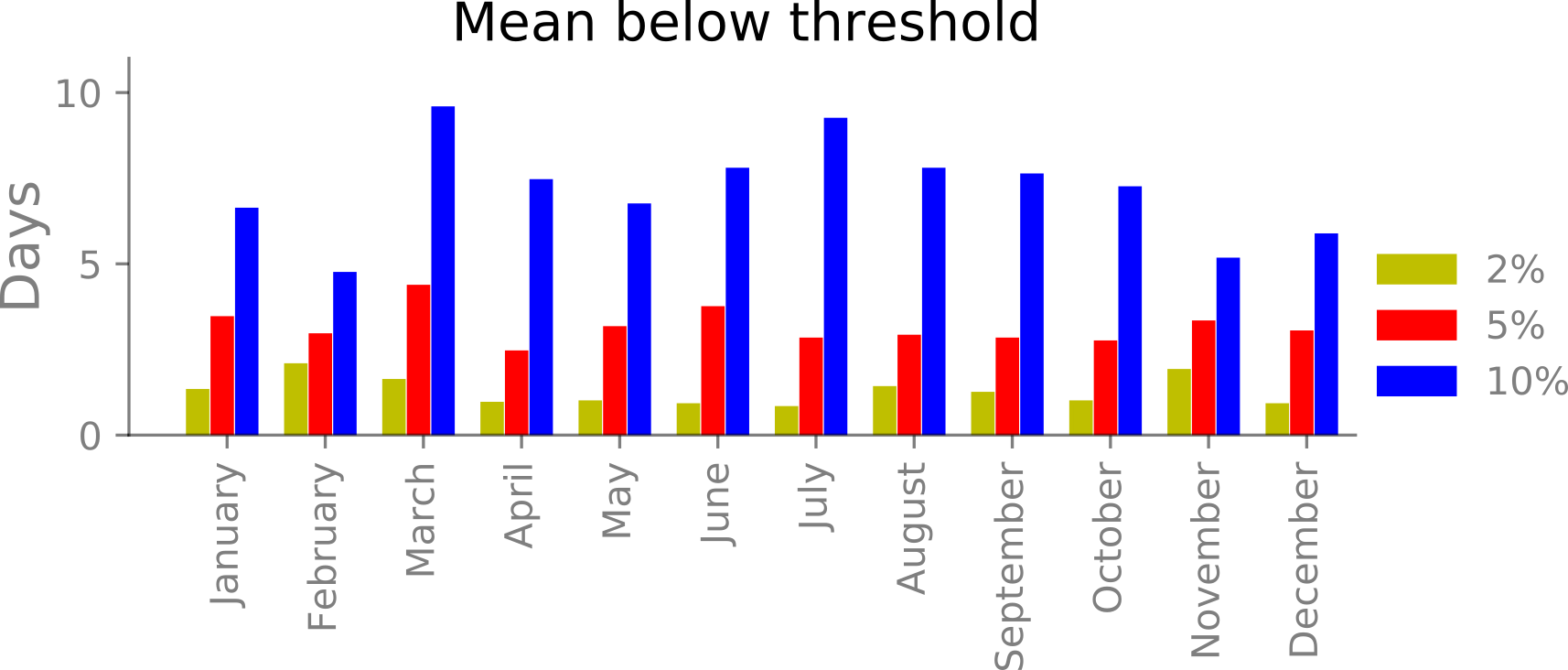}\\
\protect\caption{\label{fig: most extreme month}Most extreme LWP events per month. The vertical axis shows the duration of the longest event of all respective months for the three capacity factor thresholds.}
\end{figure}

We next look at the most extreme LWP event in a monthly perspective, irrespective of the year in which these occur (Figure~\ref{fig: most extreme month}). The most extreme events for the 10\% threshold occur in March for both definitions. This is the 1985 event discussed above, with durations of nearly four (CBT) or nearly ten consecutive days (MBT).

Considering all thresholds and both LWP definitions, there is no clear trend of the most extreme monthly LWP events. That is, substantial extreme events may occur throughout the year, and also in winter months. This contrasts the previous finding that the frequency of LWP events is generally much higher in summer than in winter, as shown in Section \ref{subsec: Seasonal distribution}. Under CBT, the most extreme events in each of the winter months are even longer than those in summer months for the 10\% capacity threshold. This finding is, however, not confirmed under the MBT definition.

\subsection{Spatial distribution of wind power during most extreme LWP event\label{subsec: Spatial distribution}}

\begin{figure}[h]
\centering{}
\includegraphics[width=0.45\textwidth]{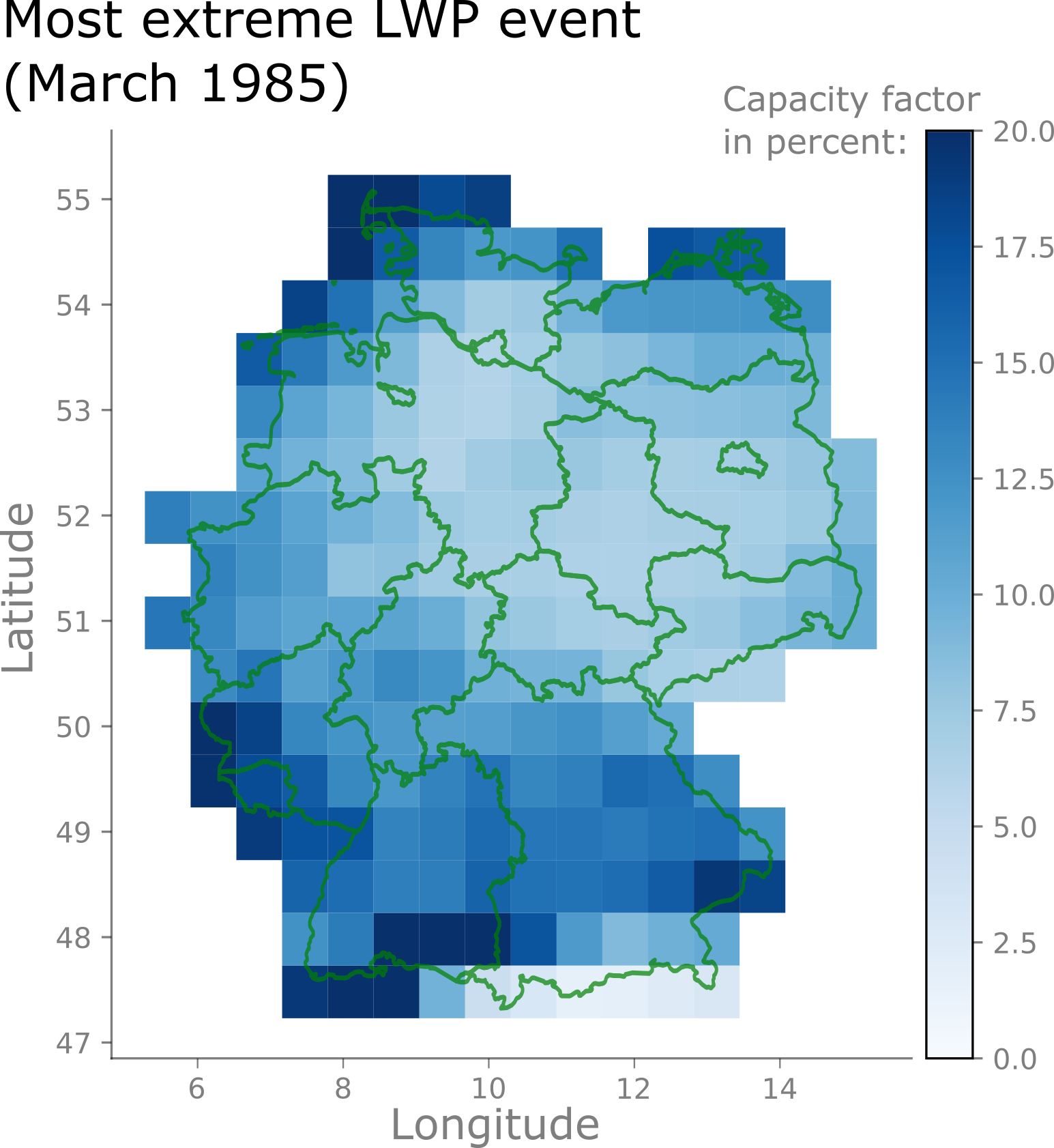} \includegraphics[width=0.45\textwidth]{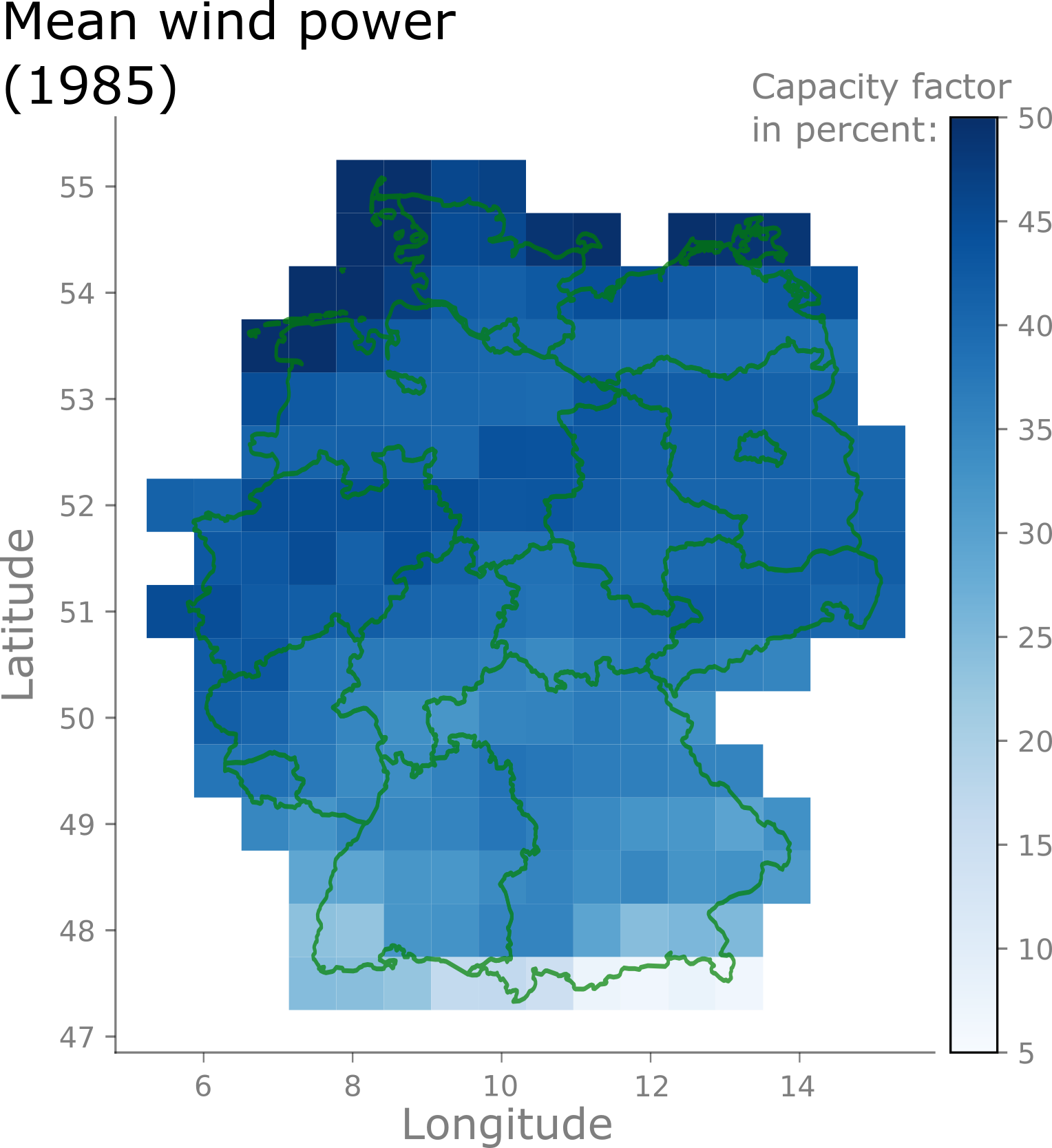}
\protect\caption{\label{fig: spatial distribution most extreme event}
Spatial distribution of wind power.
Left: Average wind power during most extreme LWP event (10\% capacity factor, MBT) in dataset in March 1985 (Scale: From 0\% to 20\% of mean capacity factors). 
Right: Mean wind power in the entire year 1985 (Scale: From 5\% to 50\% of mean capacity factors).}
\end{figure}

To also explore the spatial dimension of LWP events, we compare the distribution of capacity factors during the most extreme LWP of 1985 to the distribution of annual mean capacity factors in the same year (Figure~\ref{fig: spatial distribution most extreme event}). 

The spatial pattern of annual mean capacity factors (Figure~\ref{fig: spatial distribution most extreme event}, right panel) largely resembles that of average wind speeds in Germany (Figure~\ref{fig: Wind speed zones}).
Mean capacity factors are generally higher in Northern than in Southern Germany. They are highest close to the Northern and the Baltic Sea, and lowest in the southern Alpine region. 

The spatial pattern of mean capacity factors during the most extreme LWP event (Figure~\ref{fig: spatial distribution most extreme event}, left panel) substantially deviates from the distribution of the means. In particular, capacity factors of the north-eastern region and parts of the northern region are relatively low. The respective spatial distributions of capacity factors for other thresholds under both the CBT and MBT definitions of the same event also show substantial deviations from annual means.

Accordingly, the spatial distribution of capacity factors during extreme LWP events does not necessarily correspond to the annual mean pattern. This indicates that low-wind events can be very pronounced even in regions with very good average wind resources.

\section{Conclusions\label{sec: Conclusion}}

We analyze the seasonal distribution, frequency and magnitude of onshore low-wind-power events in Germany, as well as spatial aspects of the most extreme events, based on MERRA-2 reanalysis data and open software. We propose and evaluate two definitions of low-wind-power events for three capacity factor thresholds.

We synthesize three key results from the analysis. First, LWP events are generally most frequent in summer and least frequent in winter. Nonetheless, substantial events occur in all months of the year, and also in winter. The most persistent LWP event in the dataset occurred in March.

Second, while short events with a duration of up to around half a day are relatively frequent, very long events are much rarer.\footnote{\cite{Weber_2019} argue that low-wind event statistics do not follow a simple exponential distribution, but have ``heavy tails'', i.e.~the probability decreases rather slowly with increasing duration.} Every year, the German energy system has to deal with a period of around five consecutive days during which average wind power generation is below 10\% of the installed capacity. Every ten years, a respective period of nearly eight days is to be expected. Looking only at winter months, the durations of these expected events decrease to less than three days every winter and less than five days every tenth winter. The most persistent low-wind event in the entire dataset has a duration of nearly ten consecutive days of average wind power generation below a 10\% capacity factor.

Third, the spatial pattern of LWP events may be very different from the one of average wind power resources. During the most persistent LWP event, we find average generation to be particularly low in several regions which have some of the best wind resources.

We conclude that energy modeling studies that only consider one historic weather year are likely to substantially underestimate the occurrence of low-wind-power events and related system implications. In particular, analyses with an energy system planning perspective should take less frequent LWP events into account, e.g.,~the discussed events with a return period of ten years, or even the most extreme event identified here. This is particularly important when the complementary role of other variable and dispatchable generators, energy storage, or demand-side measures in highly-renewable energy systems is to be explored.\footnote{This is demonstrated, for example, by \cite{Schill_2018} in an analysis of storage requirements for renewable energy integration in a sensitivity analysis with one artificial no-wind week.}

Further, analyses dealing with the pros and cons of either more decentralized or more centralized renewable energy systems should consider the spatial dimension of LWP events. Although not in the focus of our analysis, our results indicate that LWP events are more pronounced for smaller geographic areas.

From an energy policy perspective, our findings on LWP events occurring in winter may be most relevant. Our analysis indicates that concerns about frequent and persistent LWP events in German winters appear to be overrated, considering that the longest event with an average capacity factor below 10\% and a ten-year return period in winter has a duration of less than five days. We further recommend that policy makers or regulators develop a proper definition of the \textit{Dunkelflaute} term, which currently appears to be used in a rather qualitative way. Our two definitions of LWP events proposed here may be useful in this context.

While our analysis deliberately focuses on LWP events of onshore wind power in Germany, we see an avenue for future research that would ideally combine the analysis of low production periods of onshore and offshore wind power as well as solar PV with time series of load, while expanding the geographic focus beyond Germany. The open-source provision of the tool used for the present analysis may be a useful starting point for such research.

\section*{Acknowledgment\label{sec: Acknowledgment}}
This analysis is a result of the research project P2X, funded by the German Federal Ministry of Education and Research (FKZ 03SFK2B1). Wolf-Peter Schill carried out parts of the work during a research stay at the University of Melbourne. Nils Ohlendorf mainly worked on this project while employed at DIW Berlin, and partly also after being employed at MCC. We thank the participants of the DIW Sustainability Cluster Seminar in April 2017, Strommarkttreffen Berlin November 2017, IAEE International Conference Groningen 2018 and Enerday Dresden 2018 for valuable comments on earlier drafts.

\section*{Data availability statement}
The data that support the findings of this study have been created with software that is openly available under an MIT license at https://doi.org/10.5281/zenodo.3694373.

\bibliography{mybibfile}

\begin{thebibliography}{46}
\expandafter\ifx\csname natexlab\endcsname\relax\def\natexlab#1{#1}\fi
\providecommand{\url}[1]{\texttt{#1}}
\providecommand{\href}[2]{#2}
\providecommand{\path}[1]{#1}
\providecommand{\DOIprefix}{doi:}
\providecommand{\ArXivprefix}{arXiv:}
\providecommand{\URLprefix}{URL: }
\providecommand{\Pubmedprefix}{pmid:}
\providecommand{\doi}[1]{\href{http://dx.doi.org/#1}{\path{#1}}}
\providecommand{\Pubmed}[1]{\href{pmid:#1}{\path{#1}}}
\providecommand{\bibinfo}[2]{#2}
\ifx\xfnm\relax \def\xfnm[#1]{\unskip,\space#1}\fi
\bibitem[{Archer and Jacobson(2007)}]{Archer_2007}
\bibinfo{author}{Archer, C.L.}, \bibinfo{author}{Jacobson, M.Z.},
  \bibinfo{year}{2007}.
\newblock \bibinfo{title}{Supplying baseload power and reducing transmission
  requirements by interconnecting wind farms}.
\newblock \bibinfo{journal}{Journal of Applied Meteorology and Climatology}
  \bibinfo{volume}{46}, \bibinfo{pages}{1701--1717}.
\newblock \DOIprefix\doi{10.1175/2007JAMC1538.1}.
\bibitem[{BMWi(2019)}]{BMWi_2019}
\bibinfo{author}{BMWi}, \bibinfo{year}{2019}.
\newblock \bibinfo{title}{{Zeitreihen zur Entwicklung der erneuerbaren Energien
  in Deutschland}}.
\newblock \bibinfo{howpublished}{Bundesministerium f{\"u}r Wirtschaft und
  Energie (Federal Ministry for Economic Affairs and Energy)}.
\newblock \URLprefix
  \url{https://www.erneuerbare-energien.de/EE/Redaktion/DE/Downloads/zeitreihen-zur-entwicklung-der-erneuerbaren-energien-in-deutschland-1990-2018.pdf}.
\bibitem[{Bosilovich et~al.(2016)Bosilovich, Lucchesi and
  Suarez}]{Bosilovich_2016}
\bibinfo{author}{Bosilovich, M.G.}, \bibinfo{author}{Lucchesi, R.},
  \bibinfo{author}{Suarez, M.}, \bibinfo{year}{2016}.
\newblock \bibinfo{title}{MERRA-2: File Specification}.
\newblock \bibinfo{type}{GMAO Office Note No. 9 (Version 1.1)}. NASA Global
  Modeling and Assimilation Office.
\newblock \URLprefix
  \url{https://gmao.gsfc.nasa.gov/pubs/docs/Bosilovich785.pdf}.
\bibitem[{Bundesregierung(2019)}]{Bundesregierung_2019}
\bibinfo{author}{Bundesregierung}, \bibinfo{year}{2019}.
\newblock \bibinfo{title}{{Klimaschutzprogramm 2030 der Bundesregierung zur
  Umsetzung des Klimaschutzplans 2050}}.
\newblock \bibinfo{howpublished}{German Federal Government}.
\newblock \URLprefix
  \url{https://www.bundesregierung.de/resource/blob/975226/1679914/e01d6bd855f09bf05cf7498e06d0a3ff/2019-10-09-klima-massnahmen-data.pdf}.
\bibitem[{Cannon et~al.(2015)Cannon, Brayshaw, Methven, Coker and
  Lenaghan}]{Cannon_2015}
\bibinfo{author}{Cannon, D.}, \bibinfo{author}{Brayshaw, D.},
  \bibinfo{author}{Methven, J.}, \bibinfo{author}{Coker, P.},
  \bibinfo{author}{Lenaghan, D.}, \bibinfo{year}{2015}.
\newblock \bibinfo{title}{Using reanalysis data to quantify extreme wind power
  generation statistics: A 33 year case study in {Great Britain}}.
\newblock \bibinfo{journal}{Renewable Energy} \bibinfo{volume}{75},
  \bibinfo{pages}{767 -- 778}.
\newblock \DOIprefix\doi{10.1016/j.renene.2014.10.024}.
\bibitem[{Carvalho et~al.(2014)Carvalho, Rocha, G\'{o}mez-Gesteira and
  Santos}]{Carvalho_2014}
\bibinfo{author}{Carvalho, D.}, \bibinfo{author}{Rocha, A.},
  \bibinfo{author}{G\'{o}mez-Gesteira, M.}, \bibinfo{author}{Santos, C.S.},
  \bibinfo{year}{2014}.
\newblock \bibinfo{title}{{WRF} wind simulation and wind energy production
  estimates forced by different reanalyses: Comparison with observed data for
  {Portugal}}.
\newblock \bibinfo{journal}{Applied Energy} \bibinfo{volume}{117},
  \bibinfo{pages}{116 -- 126}.
\newblock \DOIprefix\doi{10.1016/j.apenergy.2013.12.001}.
\bibitem[{Collins et~al.(2018)Collins, Deane, \'{O}~Gallach\'{o}ir, Pfenninger
  and Staffell}]{Collins_2018}
\bibinfo{author}{Collins, S.}, \bibinfo{author}{Deane, P.},
  \bibinfo{author}{\'{O}~Gallach\'{o}ir, B.}, \bibinfo{author}{Pfenninger, S.},
  \bibinfo{author}{Staffell, I.}, \bibinfo{year}{2018}.
\newblock \bibinfo{title}{Impacts of inter-annual wind and solar variations on
  the {European} power system}.
\newblock \bibinfo{journal}{Joule} \bibinfo{volume}{2},
  \bibinfo{pages}{2076--2090}.
\newblock \DOIprefix\doi{10.1016/j.joule.2018.06.020}.
\bibitem[{de~Coninck et~al.(2018)de~Coninck, Revi, Babiker, Bertoldi,
  Buckeridge, Cartwright, Dong, Ford, Fuss, Hourcade, Ley, Mechler, Newman,
  Revokatova, Schultz, Steg and Sugiyama}]{de_coninck_2018}
\bibinfo{author}{de~Coninck, H.}, \bibinfo{author}{Revi, A.},
  \bibinfo{author}{Babiker, M.}, \bibinfo{author}{Bertoldi, P.},
  \bibinfo{author}{Buckeridge, M.}, \bibinfo{author}{Cartwright, A.},
  \bibinfo{author}{Dong, W.}, \bibinfo{author}{Ford, J.},
  \bibinfo{author}{Fuss, S.}, \bibinfo{author}{Hourcade, J.C.},
  \bibinfo{author}{Ley, D.}, \bibinfo{author}{Mechler, R.},
  \bibinfo{author}{Newman, P.}, \bibinfo{author}{Revokatova, A.},
  \bibinfo{author}{Schultz, S.}, \bibinfo{author}{Steg, L.},
  \bibinfo{author}{Sugiyama, T.}, \bibinfo{year}{2018}.
\newblock \bibinfo{title}{Strengthening and implementing the global response},
  in: \bibinfo{booktitle}{Global {Warming} of 1.5$^{\circ}${C}. {An} {IPCC}
  {Special} {Report} on the impacts of global warming of 1.5$^{\circ}${C} above
  pre-industrial levels and related global greenhouse gas emission pathways, in
  the context of strengthening the global response to the threat of climate
  change, sustainable development, and efforts to eradicate poverty}.
\newblock \URLprefix
  \url{https://www.ipcc.ch/site/assets/uploads/sites/2/2019/05/SR15_Chapter4_Low_Res.pdf}.
\bibitem[{Decker et~al.(2012)Decker, Brunke, Wang, Sakaguchi, Zeng and
  Bosilovich}]{Decker_2012}
\bibinfo{author}{Decker, M.}, \bibinfo{author}{Brunke, M.A.},
  \bibinfo{author}{Wang, Z.}, \bibinfo{author}{Sakaguchi, K.},
  \bibinfo{author}{Zeng, X.}, \bibinfo{author}{Bosilovich, M.G.},
  \bibinfo{year}{2012}.
\newblock \bibinfo{title}{{Evaluation of the Reanalysis Products from GSFC,
  NCEP, and ECMWF Using Flux Tower Observations}}.
\newblock \bibinfo{journal}{Journal of Climate} \bibinfo{volume}{25},
  \bibinfo{pages}{1916--1944}.
\newblock \DOIprefix\doi{10.1175/JCLI-D-11-00004.1}.
\bibitem[{{Deutscher Bundestag}(2019a)}]{Bundestag_Plenarprotokoll_2019}
\bibinfo{editor}{{Deutscher Bundestag}} (Ed.), \bibinfo{year}{2019}a.
\newblock \bibinfo{title}{Plenarprotokoll 19/98 Stenografischer Bericht 98.
  Sitzung}. Plenarprotokoll 19/98.
\newblock \URLprefix \url{http://dip21.bundestag.de/dip21/btp/19/19098.pdf}.
  \bibinfo{note}{09.05.2019}.
\bibitem[{{Deutscher Bundestag}(2019b)}]{Bundestag_Unterrichtung_2019}
\bibinfo{editor}{{Deutscher Bundestag}} (Ed.), \bibinfo{year}{2019}b.
\newblock \bibinfo{title}{Unterrichtung durch die Bundesregierung Zweiter
  Fortschrittsbericht zur Energiewende 2019. Drucksache 19/10760}. Drucksache
  19/10760 19. Wahlperiode.
\newblock \URLprefix
  \url{http://dip21.bundestag.de/dip21/btd/19/107/1910760.pdf}.
  \bibinfo{note}{07.06.2019}.
\bibitem[{Gelaro et~al.(2017)Gelaro, McCarty, Suárez, Todling, Molod, Takacs,
  Randles, Darmenov, Bosilovich, Reichle, Wargan, Coy, Cullather, Draper,
  Akella, Buchard, Conaty, da~Silva, Gu, Kim, Koster, Lucchesi, Merkova,
  Nielsen, Partyka, Pawson, Putman, Rienecker, Schubert, Sienkiewicz and
  Zhao}]{Gelaro_2017}
\bibinfo{author}{Gelaro, R.}, \bibinfo{author}{McCarty, W.},
  \bibinfo{author}{Suárez, M.J.}, \bibinfo{author}{Todling, R.},
  \bibinfo{author}{Molod, A.}, \bibinfo{author}{Takacs, L.},
  \bibinfo{author}{Randles, C.A.}, \bibinfo{author}{Darmenov, A.},
  \bibinfo{author}{Bosilovich, M.G.}, \bibinfo{author}{Reichle, R.},
  \bibinfo{author}{Wargan, K.}, \bibinfo{author}{Coy, L.},
  \bibinfo{author}{Cullather, R.}, \bibinfo{author}{Draper, C.},
  \bibinfo{author}{Akella, S.}, \bibinfo{author}{Buchard, V.},
  \bibinfo{author}{Conaty, A.}, \bibinfo{author}{da~Silva, A.M.},
  \bibinfo{author}{Gu, W.}, \bibinfo{author}{Kim, G.K.},
  \bibinfo{author}{Koster, R.}, \bibinfo{author}{Lucchesi, R.},
  \bibinfo{author}{Merkova, D.}, \bibinfo{author}{Nielsen, J.E.},
  \bibinfo{author}{Partyka, G.}, \bibinfo{author}{Pawson, S.},
  \bibinfo{author}{Putman, W.}, \bibinfo{author}{Rienecker, M.},
  \bibinfo{author}{Schubert, S.D.}, \bibinfo{author}{Sienkiewicz, M.},
  \bibinfo{author}{Zhao, B.}, \bibinfo{year}{2017}.
\newblock \bibinfo{title}{{The Modern-Era Retrospective Analysis for Research
  and Applications, Version 2 (MERRA-2)}}.
\newblock \bibinfo{journal}{Journal of Climate} \bibinfo{volume}{30},
  \bibinfo{pages}{5419--5454}.
\newblock \DOIprefix\doi{10.1175/JCLI-D-16-0758.1}.
\bibitem[{Germer and Kleidon(2019)}]{Germer_2019}
\bibinfo{author}{Germer, S.}, \bibinfo{author}{Kleidon, A.},
  \bibinfo{year}{2019}.
\newblock \bibinfo{title}{Have wind turbines in germany generated electricity
  as would be expected from the prevailing wind conditions in 2000-2014?}
\newblock \bibinfo{journal}{PLOS ONE} \bibinfo{volume}{14},
  \bibinfo{pages}{1--16}.
\newblock \DOIprefix\doi{10.1371/journal.pone.0211028}.
\bibitem[{Gonz\'{a}lez-Aparicio et~al.(2017)Gonz\'{a}lez-Aparicio, Monforti,
  Volker, Zucker, Careri, Huld and Badger}]{Gonzalez_2017}
\bibinfo{author}{Gonz\'{a}lez-Aparicio, I.}, \bibinfo{author}{Monforti, F.},
  \bibinfo{author}{Volker, P.}, \bibinfo{author}{Zucker, A.},
  \bibinfo{author}{Careri, F.}, \bibinfo{author}{Huld, T.},
  \bibinfo{author}{Badger, J.}, \bibinfo{year}{2017}.
\newblock \bibinfo{title}{Simulating {European} wind power generation applying
  statistical downscaling to reanalysis data}.
\newblock \bibinfo{journal}{Applied Energy} \bibinfo{volume}{199},
  \bibinfo{pages}{155 -- 168}.
\newblock \DOIprefix\doi{10.1016/j.apenergy.2017.04.066}.
\bibitem[{Grams et~al.(2017)Grams, Beerli, Pfenninger, Staffell and
  Wernli}]{Grams_2017}
\bibinfo{author}{Grams, C.M.}, \bibinfo{author}{Beerli, R.},
  \bibinfo{author}{Pfenninger, S.}, \bibinfo{author}{Staffell, I.},
  \bibinfo{author}{Wernli, H.}, \bibinfo{year}{2017}.
\newblock \bibinfo{title}{Balancing {E}urope's wind-power output through
  spatial deployment informed by weather regimes}.
\newblock \bibinfo{journal}{Nature Climate Change} \bibinfo{volume}{7},
  \bibinfo{pages}{557--562}.
\newblock \DOIprefix\doi{10.1038/nclimate3338}.
\bibitem[{Handschy et~al.(2017)Handschy, Rose and Apt}]{Handschy_2017}
\bibinfo{author}{Handschy, M.A.}, \bibinfo{author}{Rose, S.},
  \bibinfo{author}{Apt, J.}, \bibinfo{year}{2017}.
\newblock \bibinfo{title}{Is it always windy somewhere? {Occurrence} of
  low-wind-power events over large areas}.
\newblock \bibinfo{journal}{Renewable Energy} \bibinfo{volume}{101},
  \bibinfo{pages}{1124 -- 1130}.
\newblock \DOIprefix\doi{10.1016/j.renene.2016.10.004}.
\bibitem[{{Hans Ertel Zentrum}(2019)}]{Hans-Ertel-Zentrum_2019}
\bibinfo{author}{{Hans Ertel Zentrum}}, \bibinfo{year}{2019}.
\newblock \bibinfo{title}{Cosmo regional reanalysis}.
\newblock \bibinfo{howpublished}{Universit\"{a}t Bonn and Deutscher
  Wetterdienst}.
\newblock \URLprefix \url{https://reanalysis.meteo.uni-bonn.de/?Overview}.
\bibitem[{Kahn(1979)}]{Kahn_1979}
\bibinfo{author}{Kahn, E.}, \bibinfo{year}{1979}.
\newblock \bibinfo{title}{The reliability of distributed wind generators}.
\newblock \bibinfo{journal}{Electric Power Systems Research}
  \bibinfo{volume}{2}, \bibinfo{pages}{1 -- 14}.
\newblock \DOIprefix\doi{10.1016/0378-7796(79)90021-X}.
\bibitem[{Kruyt et~al.(2017)Kruyt, Lehning and Kahl}]{Kruyt_2017}
\bibinfo{author}{Kruyt, B.}, \bibinfo{author}{Lehning, M.},
  \bibinfo{author}{Kahl, A.}, \bibinfo{year}{2017}.
\newblock \bibinfo{title}{Potential contributions of wind power to a stable and
  highly renewable {Swiss} power supply}.
\newblock \bibinfo{journal}{Applied Energy} \bibinfo{volume}{192},
  \bibinfo{pages}{1 -- 11}.
\newblock \DOIprefix\doi{10.1016/j.apenergy.2017.01.085}.
\bibitem[{Kumler et~al.(2019)Kumler, Carre{\~{n}}o, Craig, Hodge, Cole and
  Brancucci}]{Kumler_2019}
\bibinfo{author}{Kumler, A.}, \bibinfo{author}{Carre{\~{n}}o, I.L.},
  \bibinfo{author}{Craig, M.T.}, \bibinfo{author}{Hodge, B.M.},
  \bibinfo{author}{Cole, W.}, \bibinfo{author}{Brancucci, C.},
  \bibinfo{year}{2019}.
\newblock \bibinfo{title}{Inter-annual variability of wind and solar
  electricity generation and capacity values in {Texas}}.
\newblock \bibinfo{journal}{Environmental Research Letters}
  \bibinfo{volume}{14}, \bibinfo{pages}{044032}.
\newblock \DOIprefix\doi{10.1088/1748-9326/aaf935}.
\bibitem[{Leahy and McKeogh(2013)}]{Leahy_2013}
\bibinfo{author}{Leahy, P.G.}, \bibinfo{author}{McKeogh, E.J.},
  \bibinfo{year}{2013}.
\newblock \bibinfo{title}{Persistence of low wind speed conditions and
  implications for wind power variability}.
\newblock \bibinfo{journal}{Wind Energy} \bibinfo{volume}{16},
  \bibinfo{pages}{575--586}.
\newblock \DOIprefix\doi{10.1002/we.1509}.
\bibitem[{Lil\'{e}o et~al.(2013)Lil\'{e}o, Berge, Undheim, Klinkert and
  Bredesen}]{Lileo_2013}
\bibinfo{author}{Lil\'{e}o, S.}, \bibinfo{author}{Berge, E.},
  \bibinfo{author}{Undheim, O.}, \bibinfo{author}{Klinkert, R.},
  \bibinfo{author}{Bredesen, R.E.}, \bibinfo{year}{2013}.
\newblock \bibinfo{title}{Long-term correction of wind measurements.
  state-of-the-art, guidelines and future work}.
\newblock \bibinfo{journal}{Complexity} \bibinfo{volume}{1},
  \bibinfo{pages}{2--3}.
\bibitem[{Lil\'{e}o and Petrik(2011)}]{Lileo_2011}
\bibinfo{author}{Lil\'{e}o, S.}, \bibinfo{author}{Petrik, O.},
  \bibinfo{year}{2011}.
\newblock \bibinfo{title}{{Investigation on the use of NCEP/NCAR, MERRA and
  NCEP/CFSR reanalysis data in wind resource analysis}}, in:
  \bibinfo{booktitle}{European Wind Energy Conference and Exhibition 2011, EWEC
  2011}.
\bibitem[{L{\"u}ers(2016)}]{Deutsche_Windguard_2016}
\bibinfo{author}{L{\"u}ers, S.}, \bibinfo{year}{2016}.
\newblock \bibinfo{title}{Status des Windenergieausbaus an Land in Deutschland
  - Zusätzliche Auswertungen und Daten für das Jahr 2015}.
\newblock \bibinfo{type}{Technical Report}. Deutsche WindGuard.
  \bibinfo{address}{Varel}.
\newblock \URLprefix
  \url{https://www.windguard.de/veroeffentlichungen.html?file=files/cto_layout/img/unternehmen/veroeffentlichungen/2016/Status%20des%20Windenergieausbaus%20an%20Land%20in%20Deutschland%20-%20Zus%C3%A4tzliche%20Auswertungen%20und%20Daten%20f%C3%BCr%20das%20Jahr%202015.pdf}.
\bibitem[{Moemken et~al.(2018)Moemken, Reyers, Feldmann and
  Pinto}]{Moemken_2018}
\bibinfo{author}{Moemken, J.}, \bibinfo{author}{Reyers, M.},
  \bibinfo{author}{Feldmann, H.}, \bibinfo{author}{Pinto, J.G.},
  \bibinfo{year}{2018}.
\newblock \bibinfo{title}{Future changes of wind speed and wind energy
  potentials in {EURO-CORDEX} ensemble simulations}.
\newblock \bibinfo{journal}{Journal of Geophysical Research: Atmospheres}
  \bibinfo{volume}{123}, \bibinfo{pages}{6373--6389}.
\newblock \DOIprefix\doi{10.1029/2018JD028473}.
\bibitem[{Molod et~al.(2015)Molod, Takacs, Suarez and Bacmeister}]{Molod_2015}
\bibinfo{author}{Molod, A.}, \bibinfo{author}{Takacs, L.},
  \bibinfo{author}{Suarez, M.}, \bibinfo{author}{Bacmeister, J.},
  \bibinfo{year}{2015}.
\newblock \bibinfo{title}{{Development of the GEOS-5 atmospheric general
  circulation model: evolution from MERRA to MERRA2}}.
\newblock \bibinfo{journal}{Geoscientific Model Development}
  \bibinfo{volume}{8}, \bibinfo{pages}{1339--1356}.
\newblock \DOIprefix\doi{10.5194/gmd-8-1339-2015}.
\bibitem[{Ohlendorf(2020)}]{Ohlendorf_2020}
\bibinfo{author}{Ohlendorf, N.}, \bibinfo{year}{2020}.
\newblock \bibinfo{title}{{Source code for ``Frequency and persistence of
  low-wind-power events in Germany''}}.
\newblock \bibinfo{howpublished}{Zenodo}.
\newblock \DOIprefix\doi{10.5281/zenodo.3694374}.
\bibitem[{Olauson and Bergkvist(2015)}]{Olauson_2015}
\bibinfo{author}{Olauson, J.}, \bibinfo{author}{Bergkvist, M.},
  \bibinfo{year}{2015}.
\newblock \bibinfo{title}{{Modelling the Swedish wind power production using
  MERRA reanalysis data}}.
\newblock \bibinfo{journal}{Renewable Energy} \bibinfo{volume}{76},
  \bibinfo{pages}{717 -- 725}.
\newblock \DOIprefix\doi{10.1016/j.renene.2014.11.085}.
\bibitem[{{Open Power System Data}(2017)}]{OPSD_2017}
\bibinfo{author}{{Open Power System Data}}, \bibinfo{year}{2017}.
\newblock \bibinfo{title}{Data package renewable power plants}.
\newblock \URLprefix
  \url{https://data.open-power-system-data.org/renewable_power_plants/2017-02-16/}.
  \bibinfo{note}{version 2017-02-16}.
\bibitem[{Patlakas et~al.(2017)Patlakas, Galanis, Diamantis and
  Kallos}]{Patlakas_2017}
\bibinfo{author}{Patlakas, P.}, \bibinfo{author}{Galanis, G.},
  \bibinfo{author}{Diamantis, D.}, \bibinfo{author}{Kallos, G.},
  \bibinfo{year}{2017}.
\newblock \bibinfo{title}{Low wind speed events: persistence and frequency}.
\newblock \bibinfo{journal}{Wind Energy} \bibinfo{volume}{20},
  \bibinfo{pages}{1033--1047}.
\newblock \DOIprefix\doi{10.1002/we.2078}.
\bibitem[{Raynaud et~al.(2018)Raynaud, Hingray, Fran\c{c}ois and
  Creutin}]{Raynaud_2018}
\bibinfo{author}{Raynaud, D.}, \bibinfo{author}{Hingray, B.},
  \bibinfo{author}{Fran\c{c}ois, B.}, \bibinfo{author}{Creutin, J.},
  \bibinfo{year}{2018}.
\newblock \bibinfo{title}{Energy droughts from variable renewable energy
  sources in {European} climates}.
\newblock \bibinfo{journal}{Renewable Energy} \bibinfo{volume}{125},
  \bibinfo{pages}{578 -- 589}.
\newblock \DOIprefix\doi{10.1016/j.renene.2018.02.130}.
\bibitem[{Rose and Apt(2015)}]{Rose_2015}
\bibinfo{author}{Rose, S.}, \bibinfo{author}{Apt, J.}, \bibinfo{year}{2015}.
\newblock \bibinfo{title}{What can reanalysis data tell us about wind power?}
\newblock \bibinfo{journal}{Renewable Energy} \bibinfo{volume}{83},
  \bibinfo{pages}{963 -- 969}.
\newblock \DOIprefix\doi{10.1016/j.renene.2015.05.027}.
\bibitem[{Santos-Alamillos et~al.(2017)Santos-Alamillos, Brayshaw, Methven,
  Thomaidis, Ruiz-Arias and Pozo-V{\'{a}}zquez}]{Santos_Alamillos_2017}
\bibinfo{author}{Santos-Alamillos, F.J.}, \bibinfo{author}{Brayshaw, D.J.},
  \bibinfo{author}{Methven, J.}, \bibinfo{author}{Thomaidis, N.S.},
  \bibinfo{author}{Ruiz-Arias, J.A.}, \bibinfo{author}{Pozo-V{\'{a}}zquez, D.},
  \bibinfo{year}{2017}.
\newblock \bibinfo{title}{Exploring the meteorological potential for planning a
  high performance {European} electricity super-grid: optimal power capacity
  distribution among countries}.
\newblock \bibinfo{journal}{Environmental Research Letters}
  \bibinfo{volume}{12}, \bibinfo{pages}{114030}.
\newblock \DOIprefix\doi{10.1088/1748-9326/aa8f18}.
\bibitem[{Schill and Zerrahn(2018)}]{Schill_2018}
\bibinfo{author}{Schill, W.P.}, \bibinfo{author}{Zerrahn, A.},
  \bibinfo{year}{2018}.
\newblock \bibinfo{title}{Long-run power storage requirements for high shares
  of renewables: Results and sensitivities}.
\newblock \bibinfo{journal}{Renewable and Sustainable Energy Reviews}
  \bibinfo{volume}{83}, \bibinfo{pages}{156 -- 171}.
\newblock \DOIprefix\doi{10.1016/j.rser.2017.05.205}.
\bibitem[{Schlott et~al.(2018)Schlott, Kies, Brown, Schramm and
  Greiner}]{Schlott_2018}
\bibinfo{author}{Schlott, M.}, \bibinfo{author}{Kies, A.},
  \bibinfo{author}{Brown, T.}, \bibinfo{author}{Schramm, S.},
  \bibinfo{author}{Greiner, M.}, \bibinfo{year}{2018}.
\newblock \bibinfo{title}{The impact of climate change on a cost-optimal highly
  renewable {European} electricity network}.
\newblock \bibinfo{journal}{Applied Energy} \bibinfo{volume}{230},
  \bibinfo{pages}{1645 -- 1659}.
\newblock \DOIprefix\doi{10.1016/j.apenergy.2018.09.084}.
\bibitem[{Shaner et~al.(2018)Shaner, Davis, Lewis and Caldeira}]{Shaner_2018}
\bibinfo{author}{Shaner, M.R.}, \bibinfo{author}{Davis, S.J.},
  \bibinfo{author}{Lewis, N.S.}, \bibinfo{author}{Caldeira, K.},
  \bibinfo{year}{2018}.
\newblock \bibinfo{title}{{Geophysical constraints on the reliability of solar
  and wind power in the United States}}.
\newblock \bibinfo{journal}{Energy and Environmental Science}
  \bibinfo{volume}{11}, \bibinfo{pages}{914--925}.
\newblock \DOIprefix\doi{10.1039/C7EE03029K}.
\bibitem[{Sharp et~al.(2015)Sharp, Dodds, Barrett and Spataru}]{Sharp_2015}
\bibinfo{author}{Sharp, E.}, \bibinfo{author}{Dodds, P.},
  \bibinfo{author}{Barrett, M.}, \bibinfo{author}{Spataru, C.},
  \bibinfo{year}{2015}.
\newblock \bibinfo{title}{{Evaluating the accuracy of CFSR reanalysis hourly
  wind speed forecasts for the UK, using in situ measurements and geographical
  information}}.
\newblock \bibinfo{journal}{Renewable Energy} \bibinfo{volume}{77},
  \bibinfo{pages}{527 -- 538}.
\newblock \DOIprefix\doi{10.1016/j.renene.2014.12.025}.
\bibitem[{Staffell and Green(2014)}]{Staffell_2014}
\bibinfo{author}{Staffell, I.}, \bibinfo{author}{Green, R.},
  \bibinfo{year}{2014}.
\newblock \bibinfo{title}{How does wind farm performance decline with age?}
\newblock \bibinfo{journal}{Renewable Energy} \bibinfo{volume}{66},
  \bibinfo{pages}{775 -- 786}.
\newblock \DOIprefix\doi{10.1016/j.renene.2013.10.041}.
\bibitem[{Staffell and Pfenninger(2016)}]{Staffell_2016}
\bibinfo{author}{Staffell, I.}, \bibinfo{author}{Pfenninger, S.},
  \bibinfo{year}{2016}.
\newblock \bibinfo{title}{Using bias-corrected reanalysis to simulate current
  and future wind power output}.
\newblock \bibinfo{journal}{Energy} \bibinfo{volume}{114}, \bibinfo{pages}{1224
  -- 1239}.
\newblock \DOIprefix\doi{10.1016/j.energy.2016.08.068}.
\bibitem[{Tobin et~al.(2016)Tobin, Jerez, Vautard, Thais, van Meijgaard, Prein,
  D{\'{e}}qu{\'{e}}, Kotlarski, Maule, Nikulin, Noël and
  Teichmann}]{Tobin_2016}
\bibinfo{author}{Tobin, I.}, \bibinfo{author}{Jerez, S.},
  \bibinfo{author}{Vautard, R.}, \bibinfo{author}{Thais, F.},
  \bibinfo{author}{van Meijgaard, E.}, \bibinfo{author}{Prein, A.},
  \bibinfo{author}{D{\'{e}}qu{\'{e}}, M.}, \bibinfo{author}{Kotlarski, S.},
  \bibinfo{author}{Maule, C.F.}, \bibinfo{author}{Nikulin, G.},
  \bibinfo{author}{Noël, T.}, \bibinfo{author}{Teichmann, C.},
  \bibinfo{year}{2016}.
\newblock \bibinfo{title}{Climate change impacts on the power generation
  potential of a {European} mid-century wind farms scenario}.
\newblock \bibinfo{journal}{Environmental Research Letters}
  \bibinfo{volume}{11}, \bibinfo{pages}{034013}.
\newblock \DOIprefix\doi{10.1088/1748-9326/11/3/034013}.
\bibitem[{Wallasch et~al.(2015)Wallasch, L{\"u}ers and
  Rehfeldt}]{Deutsche_Windguard_2015}
\bibinfo{author}{Wallasch, A.K.}, \bibinfo{author}{L{\"u}ers, S.},
  \bibinfo{author}{Rehfeldt, K.}, \bibinfo{year}{2015}.
\newblock \bibinfo{title}{Kostensituation der Windenergie an Land in
  Deutschland - Update}.
\newblock \bibinfo{type}{Technical Report}. Deutsche WindGuard.
  \bibinfo{address}{Varel}.
\newblock \URLprefix
  \url{https://www.windguard.de/veroeffentlichungen.html?file=files/cto_layout/img/unternehmen/veroeffentlichungen/2015/Kostensituation%20der%20Windenergie%20an%20Land%20in%20Deutschland%20-%20Update.pdf}.
\bibitem[{Weber et~al.(2019)Weber, Reyers, Beck, Timme, Pinto, Witthaut and
  Sch\"{a}fer}]{Weber_2019}
\bibinfo{author}{Weber, J.}, \bibinfo{author}{Reyers, M.},
  \bibinfo{author}{Beck, C.}, \bibinfo{author}{Timme, M.},
  \bibinfo{author}{Pinto, J.G.}, \bibinfo{author}{Witthaut, D.},
  \bibinfo{author}{Sch\"{a}fer, B.}, \bibinfo{year}{2019}.
\newblock \bibinfo{title}{Wind power persistence characterized by
  superstatistics}.
\newblock \bibinfo{journal}{Scientific Reports} \bibinfo{volume}{9},
  \bibinfo{pages}{19971--}.
\newblock \DOIprefix\doi{10.1038/s41598-019-56286-1}.
\bibitem[{Wetzel(2017)}]{Welt_2017}
\bibinfo{author}{Wetzel, D.}, \bibinfo{year}{2017}.
\newblock \bibinfo{title}{{Die ,,Dunkelflaute'' bringt Deutschlands
  Stromversorgung ans Limit}}.
\newblock \bibinfo{journal}{Die Welt.} \URLprefix
  \url{https://www.welt.de/wirtschaft/article161831272/Die-Dunkelflaute-bringt-Deutschlands-Stromversorgung-ans-Limit.html}.
\bibitem[{Wetzel(2019)}]{Welt_2019}
\bibinfo{author}{Wetzel, D.}, \bibinfo{year}{2019}.
\newblock \bibinfo{title}{{In der ,,kalten Dunkelflaute'' r\"{a}cht sich die
  Energiewende}}.
\newblock \bibinfo{journal}{Die Welt.} \URLprefix
  \url{https://www.welt.de/wirtschaft/article191195983/Energiewende-Das-droht-uns-in-der-kalten-Dunkelflaute.html}.
\bibitem[{Wiese et~al.(2019)Wiese, Schlecht, Bunke, Gerbaulet, Hirth, Jahn,
  Kunz, Lorenz, Mühlenpfordt, Reimann and Schill}]{Wiese_2019}
\bibinfo{author}{Wiese, F.}, \bibinfo{author}{Schlecht, I.},
  \bibinfo{author}{Bunke, W.D.}, \bibinfo{author}{Gerbaulet, C.},
  \bibinfo{author}{Hirth, L.}, \bibinfo{author}{Jahn, M.},
  \bibinfo{author}{Kunz, F.}, \bibinfo{author}{Lorenz, C.},
  \bibinfo{author}{Mühlenpfordt, J.}, \bibinfo{author}{Reimann, J.},
  \bibinfo{author}{Schill, W.P.}, \bibinfo{year}{2019}.
\newblock \bibinfo{title}{{Open Power System Data – Frictionless data for
  electricity system modelling}}.
\newblock \bibinfo{journal}{Applied Energy} \bibinfo{volume}{236},
  \bibinfo{pages}{401 -- 409}.
\newblock \DOIprefix\doi{10.1016/j.apenergy.2018.11.097}.
\bibitem[{{Wissenschaftliche Dienste}(2019)}]{WD_2019}
\bibinfo{author}{{Wissenschaftliche Dienste}}, \bibinfo{year}{2019}.
\newblock \bibinfo{title}{{Sicherstellung der Stromversorgung bei
  Dunkelflauten}}.
\newblock \bibinfo{type}{Documentation}. Deutscher Bundestag.
\newblock \URLprefix
  \url{https://www.bundestag.de/resource/blob/627898/b65deea51fdb399e4b64f1182465658d/WD-5-167-18-pdf-data.pdf}.
  \bibinfo{note}{wD 5 - 3000 - 167/18}.

\end{thebibliography}

\newpage
\appendix

\section{Reanalysis data and its use for energy modelling}\label{appendix_reanalysis}

Reanalysis data is increasingly used for energy modelling as it provides consistent global time series of long-term atmosphere data such as wind speed, temperature and air pressure in regular spatial and temporal resolutions. The underlying global circulation models extrapolate measurement station data on wind speeds, temperature, moisture and surface pressure as well as data from satellites and precipitation measurements \citep{Decker_2012}. Several publicly available second-generation global reanalysis datasets have been released since the early 2000s. We use MERRA-2, which builds on and improves the previous MERRA dataset, using advanced models and data sources \citep{Molod_2015}.

\cite{Decker_2012} evaluate the accuracy of several reanalysis datasets (MERRA, NCEP, ERA-40, ERA-Interim, CFSR and GLDAS) using flux tower measurements in the Northern Hemisphere. Almost all products overestimate the monthly and 6-hourly wind speeds and their variability. MERRA and ERA-Interim show the lowest values root-mean-square error and bias for diurnal cycles. \cite{Sharp_2015} review other data validation studies of different reanalysis datasets. Three studies derive Pearson’s correlation coefficients for MERRA between 0.75 and 0.89 based on measurement stations in Sweden, Portugal, Norway and Denmark \citep{Lileo_2011, Lileo_2013, Carvalho_2014}. \cite{Staffell_2016} propose country-specific wind speed bias correction factors for MERRA and MERRA-2 to increase the correlation with national capacity factors. Without such correction, average capacity factors for Germany based on raw MERRA or MERRA-2 wind speeds would be overestimated. \cite{Staffell_2014} make a similar point for the UK. In contrast, \cite{Cannon_2015} do not use correction factors in a UK application. Even if MERRA wind speeds turn out to be not particularly valid for single measurement points, spatial aggregation of mean wind speed over all stations results in a correlation coefficient of 0.94. This indicates a high validity of MERRA data for large-scale wind patterns. Following \cite{Cannon_2015}, we also refrain from introducing correction factors and instead make use of the error-smoothing effect of spatial aggregation. In doing so, we also avoid model artefacts, particularly as the usefulness of correction factors has only been demonstrated for average wind speeds, but not for extreme values.

\section{Wind power turbines\label{sec: Wind_turbines}}

The low- and high-wind power curves used in our analysis are based on data of eight wind power turbines by six manufacturers, namely Nordex, Senvion, Enercon, Vestas, Gamesa and Vensys. Specifically, we use the following high-wind power turbines:
\begin{itemize}
    \item Nordex N90-2.5MW
    \item Vestas V90-2.0MW
    \item Gamesa G97-2MW
    \item Vensys 100-2.5MW
\end{itemize}
Analogously, we use following low-wind power turbines:
\begin{itemize}
    \item Nordex N131-3.3MW
    \item Senvion 3.2M122
    \item Enercon E126 EP4/4.2MW
    \item Vestas V126-3.3MW
\end{itemize}

\section{Discussion of limitations\label{sec: Discussion}}

We briefly discuss some limitations of our analysis and how these may qualitatively impact results.

First, there are general limitations of using reanalysis data which have been discussed in the literature, for example spatial biases or issues with upscaling to hub heights \citep{Sharp_2015, Olauson_2015, Rose_2015, Staffell_2016}. It is, however, not clear if there are specific distortions with respect to extreme low-wind events derived from reanalysis data. A limitation specific to the MERRA-2 dataset is the relatively coarse 50x50 km grid cell size, which insufficiently represent local impacts on wind speeds. Regional reanalysis data with more refined geographical resolutions may resolve this issue, e.g.~COSMO-REA2 with 2x2 km, or COSMO-REA6 with 6x6 km \citep{Hans-Ertel-Zentrum_2019}, yet these are only available for shorter periods of time. The global coverage of MERRA-2 further allows repeating our open-source analysis for other countries and world regions. 

Second, we use power curves of currently available wind turbines and assume hub-heights of recently constructed plants. We may thus overestimate wind power generation compared to the currently existing fleet of wind turbines in Germany, which includes many older and smaller turbines, and in turn underestimate the magnitude of current LWP events. Conversely, we may underestimate power generation of future turbines, and accordingly overestimate the magnitude of future low-wind-power events, assuming that turbine efficiency and hub height increases further, with corresponding upward shifts in the power curves. Once LWP events become more relevant for the overall energy system, this may also trigger specific technology improvements toward lower cut-in speeds and a steeper slope of the power curve on the very left-hand side. Quantifying the potentially mitigating effects of such developments on LWP periods is left for future research.

Third, we use the current spatial capacity distribution of German wind power plants for deriving an aggregated capacity factor time series. We implicitly assume that this distribution also persists in the future. In reality, a relative increase of wind power deployment at sites with lower wind resources may occur, for example in southern Germany. From the results presented in Section \ref{subsec: Seasonal distribution}, we infer that a more even spatial dispersion of wind turbines could slightly mitigate LWP events.

Next, climate change has an impact on wind speeds. Future time series of wind power capacity factors will thus differ from the historic ones investigated here. \cite{Tobin_2016} find that wind power variability in Europe may generally increase, but \cite{Schlott_2018} conclude that this has no substantial effect on optimal deployment of onshore wind power in highly renewable future scenarios. \cite{Moemken_2018} find that climate change will increase the occurrence of low wind speeds. 

Finally, the focus of this analysis is a detailed but selective investigation of onshore LWP events in Germany. This geographic focus helps to keep the analysis tractable and avoids making implicit assumptions on continental electricity transmission infrastructure. It is also relevant from an energy policy perspective, which often includes national energy security considerations. Yet expanding the geographic scope of the analysis would allow raising complementary insights on larger-scale spatial patterns of extreme LWP events. Focusing on onshore wind power, and not including other renewable energy sources such as offshore wind power and solar PV, allows for parsimonious model assumptions, and findings remain valid for any level of installed capacity. Analyses that would combine periods of low production from various renewable energy sources, and also explore their correlation with electric load, appear to be a promising field for future research. The work of \cite{Raynaud_2018}, albeit with lower temporal and spatial detail compared to our analysis, can be considered as a first step in this direction.

\end{document}